\documentclass{aa}  

\usepackage{graphicx}
\usepackage{txfonts}
\usepackage{natbib,twoopt}

\begin{document}

   \title{Exploring circumstellar effects on the lithium and calcium abundances in massive Galactic O-rich AGB stars}

   \author{V. P\'erez-Mesa\inst{1,2}, O. Zamora\inst{1,2}, D. A. Garc\'ia-Hern\'andez\inst{1,2}, Y. Osorio\inst{1,2}, T. Masseron\inst{1,2}, B. Plez\inst{3}, A. Manchado\inst{1,2,4}, A. I. Karakas\inst{5} \and M. Lugaro\inst{5,6}
          }

   \institute{Instituto de Astrof\'isica de Canarias (IAC), E-38205 La Laguna, Tenerife, Spain\\
              \email{vperezme@iac}
         \and
             Universidad de La Laguna (ULL), Departamento de Astrof\'isica, E-38206 La Laguna, Tenerife, Spain
         \and
             Laboratoire Univers et Particules de Montpellier, Universit\'e de Montpellier2, CNRS, 34095 Montpellier, France
             \and
             Consejo Superior de Investigaciones Cient\'ificas (CSIC), E-28006 Madrid, Spain
             \and
              Monash Centre for Astrophysics, School of Physics and Astronomy, Monash University, VIC3800, Australia
             \and
             Konkoly Observatory, Research Centre for Astronomy and Earth Sciences, Hungarian Academy of Sciences, 1121 Budapest, Hungary\\
             }

  \date{Received November 9, 2018; accepted February 7, 2019}

\titlerunning{Exploring circumstellar effects on the Li and Ca abundances in massive Galactic O-rich AGB stars}
\authorrunning{V. P\'erez-Mesa et al.}

% \abstract{}{}{}{}{} 
% 5 {} token are mandatory
 
  \abstract
  % context heading (optional)
  % {} leave it empty if necessary  
   {We previously explored the circumstellar effects on the Rb and Zr
abundances in massive Galactic O-rich AGB
stars. Here we are interested in clarifying the role of the extended atmosphere 
in the case of Li and Ca. Li is an important indicator of hot bottom
burning (HBB) while the total Ca abundances in these stars could be affected by
neutron captures.}
  % aims heading (mandatory)
   {We report new Li and Ca abundances in massive
Galactic O-rich AGB stars by using extended model atmospheres.
The Li abundances were previously studied with hydrostatic models, while the Ca
abundances have been determined here for the first time.}
  % methods heading (mandatory)
   {We use a modified version of Turbospectrum and
consider the presence of a gaseous circumstellar envelope and radial wind.
The Li and Ca abundances
are obtained from the 6708 $\AA$ Li I and 6463 $\AA$ Ca I resonance lines,
respectively. In addition, we study the sensitivity of the pseudo-dynamical
models to variations of the stellar and wind parameters.}
  % results heading (mandatory)
   {The Li abundances derived with the pseudo-dynamical models are
very similar to those obtained from hydrostatic models (the average difference
is 0.18 dex, $\sigma^{2}$ = 0.02), with no difference for Ca. The Li and Ca 
content in these stars is only slightly affected by the 
presence of a circumstellar envelope. We also found that the Li I and Ca I line
profiles are not very sensitive to variations of the model wind parameters.}
  % conclusions heading (optional), leave it empty if necessary 
   {The new Li abundances confirm the Li-rich (and super Li-rich)
nature of the sample stars, supporting the activation of HBB
in massive Galactic AGB stars. This is in good 
agreement with the theoretical predictions
for solar metallicity AGB models from ATON, Monash, and NuGrid/MESA  but is at
odds with the FRUITY database, which predicts no hot bottom
burning leading to the production
of Li. Most sample stars display nearly solar (within the estimated errors
and considering possible NLTE) Ca abundances that are consistent with
the available \textit{s}-process nucleosynthesis models for solar metallicity
massive AGB stars, which predict overproduction of $^{46}$Ca relatively to the 
other Ca isotope and the creation of the radiactive isotope $^{41}$Ca 
but no change in the total Ca abundance. A minority of the
sample stars seem to show a significant Ca depletion (by up to 1.0 dex). Possible
explanations are offered to explain their apparent and unexpected Ca
depletion.}


   \maketitle
\textbf{}%
%________________________________________________________________

\section{Introduction}

Stars with initial masses in the range between 0.8 and 8 M$_{\odot}$ end their
lives with a phase of strong mass loss and thermal pulses (TP) on the asymptotic
giant branch (AGB; e.g. \citealt{herwig05, karakaslattanzio14}). AGB stars
are one of the main contributors to the chemical enrichment of the interstellar
medium (ISM) of light elements (e.g. Li, C, N, F) and heavy \citep[\textit{slow}
neutron capture, \textit{s}-process; e.g.] []{busso01} elements (e.g. Rb, Zr,
Tc, etc.) and so to the chemical evolution of complex stellar systems such as
galaxies and globular clusters.

AGB stars are also an important source of dust in galaxies and the site of 
origin of the vast majority of meteoritic stardust grains \citep[e.g.][]
{hoppeott97, nittler97, lugaro17}. The low-mass AGB stars (M $<$ 3$-$4
M$_{\odot}$)  are C-rich stars (C/O $>$ 1) because $^{12}$C is produced during
the TP-AGB  phase and carried to the stellar surface via the third dredge-up
(TDU),  transforming O-rich stars into C-rich ones \citep{herwig05,
karakaslattanzio07,  lugarochieffi11}. On the other hand, the more massive AGB
stars (M $>$ 4-5  M$_{\odot}$) are O-rich (C/O $<$ 1) due to the activation of
the so-called  "hot bottom burning" (hereafter, HBB) process. HBB converts $^{12}$C
into $^{13}$C  and $^{14}$N through the CN cycle via proton captures at the base
of the convective envelope, preventing the formation of a carbon star
\citep[e.g.][]{sackmann92, mazzitelli99}. The HBB models \citep[e.g.][]{sackmann92,
mazzitelli99} predict also the production of $^{7}$Li via the  "$^{7}$Be
transport mechanism" \citep{cameron-fowler71}, where Li should be detectable at
the stellar surface regions (at least for  a short time; see below).

%The HBB process is activated when temperature reaches 40 MK at the bottom of
%the envelope, the same needed temperature to ignite the so-called "$^{7}$Be 
%transport mechanism", which predicts the production of the short-lived 
%$^{7}$Li by the chain $^{3}$He($\alpha$, $\gamma$)$^{7}$Be(e$^{-}$, $v$)$^{7}$Li 
%\citep{cameron-fowler71}. The HBB models \citep{sackmann92, dantona96, 
%mazzitelli96, mazzitelli99} predict great amounts of Li in the surface regions. 
%The models which do not considere the activation of the HBB cannot explain the
%(super-)Li-rich character of the massive AGB stars \citep{cristallo15}.
%On the other hand, in the more massive AGB stars, which experience HBB, the
%$^{22}$Ne $+$ $\alpha$ $\longrightarrow$ $^{25}$Mg $+$ n neutron source
%reaction is significantly activating. These free neutrons drive the production 
%of $^{41}$Ca, and then, the produced $^{41}$Ca is carried up to the stellar 
%surface from the inteshell region via the TDU \citep[see][]{lugaro12}.

Regarding the \textit{s}-process, the $^{13}$C($\alpha$,n)$^{16}$O reaction
operates during the  interpulse period and is the preferred neutron source in
low-mass  AGB stars \citep[e.g.][]{lambert95, abia01}. The neutrons are
captured  by iron nuclei and other heavy elements forming \textit{s}-elements
that can  later be dredged to the stellar surface 
\citep[e.g.][]{busso01, karakaslattanzio14}. Another neutron source, 
$^{22}$Ne($\alpha$,n)$^{25}$Mg, requires higher temperatures and produces higher
neutron densities (up to 10$^{13}$ n/cm$^{3}$) than the $^{13}$C($\alpha$,n)$^{16}$O 
reaction \citep[see e.g.][]{vanraai12, fishlock14}. The $^{22}$Ne($\alpha$,n)$^{25}$Mg
reaction  operates during the convective TP and dominates the production of 
\textit{s}-process elements in the more massive (M $>$ 4-5 M$_{\odot}$) AGB
stars  \citep[e.g.][]{garcia-hernandez06,garcia-hernandez09,garcia-hernandez13}.
A different \textit{s}-elements pattern is expected depending on the dominant
neutron source; in particular a higher amount of Rb compared with neighboring
elements like Zr and Sr. Interestingly, the free neutrons can drive neutron 
captures also on the light elements including the Ca isotopes. While the total 
abundance of Ca is predicted not to vary in AGB stars by more than roughly 10\% 
\citep[e.g.][]{karakaslugaro16}, the isotopic composition of Ca can be affected, 
mostly resulting in an overproduction of $^{46}$Ca relatively to the other Ca 
isotopes \citep[see also][]{wasserburg15}. Furthermore, the models predict the 
production of the radionuclide $^{41}$Ca (half life 0.1 Myr), which can also be 
carried up to the stellar surface from the intershell region via the TDU, with 
maximum $^{41}$Ca/$^{40}$Ca ratios at the stellar surface of the order of 10$^{-4}$ 
\citep[see e.g.][]{trigo-rodriguez09, lugaro12, lugaro14}. This isotope decays 
via electron captures and is also destroyed by neutron captures via 
$^{41}$Ca(n,$\alpha$)$^{38}$Ar and $^{41}$Ca(n,p)$^{41}$K. All of these interaction 
channels are uncertain \citep[see][for a discussion]{lugaro18}, so the production 
of $^{41}$Ca could in principle lead to a decrease in the total Ca abundance. 
However, the cross section of the production channel of $^{41}$Ca, the 
$^{40}$Ca(n,$\gamma$)$^{41}$Ca reaction is well determined \citep{dillmann09}
and we do not expect major changes in the model predictions if any of the input 
physics related to $^{41}$Ca will be modified.

The first photometric identification of massive O-rich AGB stars was done in
the  Magellanic Clouds (MCs) about 30 years ago \citep{wood83}. These stars were
found to be long-period variables ($\sim$500-800 days) of Mira type and enriched
in heavy neutron-rich s-process elements \citep[see][for more details]{wood83}.
Subsequent high-resolution optical spectroscopic observations of AGB stars in
both MCs (LMC and SMC) discovered that these stars are Li-rich, confirming the
activation of HBB \citep[see e.g.][]{smithlambert89,smithlambert90,plez93,smith95}. More detailed chemical
analysis show that the Li-rich HBB stars in the SMC display low C isotopic
ratios, near to the equilibrium values, as expected from HBB models; these
stars, however, are not rich in Rb but rich in other  s-process elements like Zr
and Nd \citep{plez93}. This observation suggests that these low-metallicity HBB
stars mainly produce s-process elements via the $^{13}$C neutron source
\citep[e.g.][]{abia01,garcia-hernandez09,karakas18}. More recently, candidate HBB stars
have been identified in the very low metallicity ([Fe/H] $\sim$ $-$1.6) dwarf
galaxy  IC 1613, but they are likely younger and more metal-rich than the
average IC 1613 metallicity; one of these stars displays a strong Li I
6708\AA~line and it is probably Li-rich \citep{menzies15}. In our own Galaxy,
high-resolution optical spectroscopic surveys of very luminous OH/IR stars show
that most of the stars with long periods and high OH expansion velocities are
Li-rich, which confirm them as truly massive HBB-AGB stars. The strong Rb
overabundaces coupled with mild Zr enhancements \citep{garcia-hernandez06,
garcia-hernandez07} confirm the activation of the $^{22}$Ne neutron source
in the more massive O-rich AGB stars. More recently, observations of a few
massive Galactic AGB stars at the beginning of the TP phase have confirmed that
HBB is strongly activated at the early AGB stages and that the
\textit{s}-process is dominated by the $^{22}$Ne neutron source
\citep{garcia-hernandez13}. The latter stars are super Li-rich
(log$\varepsilon$(Li) up to $\sim$4 dex) together with the lack of
\textit{s}-process elements (Rb, Zr and Tc), as predicted by the theoretical
models \citep[e.g.][]{vanraai12, karakas12}. On the other hand, the Ca
abundances have never been previously measured in massive Galactic AGB stars;
here we report for the first time the Ca abundances in a complete sample of such
stars. 

The chemical abundance analyses of the massive AGB stars of our Galaxy and the
MCs \citep[generally OH/IR stars;][]{garcia-hernandez06,garcia-hernandez07,garcia-hernandez09} were made by
using classical MARCS  hydrostatic atmospheres \citep{gustafsson08}. The
analysis confirm that the $^{22}$Ne neutron source dominates the production of
\textit{s}-elements in these stars but the theoretical models cannot explain the
extremely high Rb abundances and [Rb/Zr] ratios observed (especially in the
lower metallicity MC-AGB stars). \cite{zamora14} constructed new
pseudo-dynamical MARCS model atmospheres in which the presence of a gaseous
circumstellar envelope and radial wind are considered and applied them to a
small sample of O-rich AGB stars. The Rb abundances and [Rb/Zr] ratios obtained
by \cite{zamora14} are much lower; in better  agreement with the AGB
nucleosynthesis models. More recently, \cite{perez-mesa17} reported the
pseudo-dynamical Rb and Zr abundances in a larger sample of massive Galactic AGB
stars, previoulsy studied with hydrostatic models 
\citep[see][]{garcia-hernandez06,garcia-hernandez07},  by using the more
realistic \cite{zamora14} extended model atmospheres.  The new Rb abundances and
[Rb/Zr] ratios obtained by \cite{perez-mesa17} are much lower and in much better 
agreement with the AGB theoretical predictions, significantly resolving the mismatch between the
observations and the nucleosynthesis models, and confirming the earlier
\cite{zamora14} preliminary results on a smaller sample of massive O-rich
AGBs \citep[see][for more details]{perez-mesa17}.
In this paper, we explore the circumstellar effects on the Li and Ca
abundances by applying the \cite{zamora14} pseudo-dynamical model atmospheres to
the sample of massive Galactic AGB stars of \cite{garcia-hernandez07} and the
super  Li-rich AGBs of \cite{garcia-hernandez13}. These new Li and Ca
abundances  are then compared with several AGB nucleosynthesis theoretical
predictions: ATON \citep{ventura09}, Monash \citep{karakaslugaro16}, NuGrid/MESA
\citep{ritter18} and FRUITY \citep{cristallo15} models.

%
%__________________________________________________________________

\section{Observational data}

We have used the high S/N (at least $\sim$30$-$50 around Li I 6708 $\AA$; see
below) and high-resolution (R$\sim$50,000) optical ($\sim$4000$-$9000 $\AA$)
echelle spectra for the \cite{garcia-hernandez06} sample (15 stars) of massive
Galactic AGB stars, for which Rb and Zr abundances could be derived by
\cite{perez-mesa17} as well as for the \cite{garcia-hernandez07} subsample (12
stars) of Li-detected stars not analysed by \cite{perez-mesa17}. In addition, we
have analysed the high-quality optical echelle spectra of the three (RU Cug,
SV Cas and R Cen) massive Galactic AGB stars, two of them (SV Cas and R Cen)
super Li-rich, reported by \cite{garcia-hernandez13}. The high-resolution 
spectra were obtained using the Utrecht Echelle Spectrograph (UES) at the 4.2 m
William Herschel Telescope, the CAsegrain Echelle SPECtrograph (CASPEC) at the 
ESO 3.6 m telescope, the Tull spectrograph at the 2.7 m Harlan J. Smith (HJS)
Telescope and the UVES spectrograph at the ESO-VLT \citep[see][for more observational details]
{garcia-hernandez07, garcia-hernandez13}. Our final sample is thus
composed by 30 stars; all of them with previous hydrostatic Li abundance
determinations. It is to be noted here that our subsample of stars with previous
Rb abundance determinations slightly differs from the \cite{perez-mesa17} sample
mentioned above because the observed spectra are extremely red and the S/N
ratios achieved for a given star can strongly vary from the blue to the red
spectral regions (e.g. 10$-$20 at Ca I 6463 \AA~while $>$50$-$100 at Rb I 7800
$\AA$; see also below); i.e. six stars from \cite{perez-mesa17} display a too
low S/N at Li I 6708\AA~to estimate their Li abundances and were removed from
the present sample. 

We first carried out an exhaustive study of the Li and Ca absorption spectral
lines that can be useful for the extraction of the Li and Ca abundances in these
stars. As previouly found by \cite{garcia-hernandez07}, we find the Li I 6708
$\AA$ line to be the best one for the abundance analysis; e.g. we discarded the
subordinate and weaker Li I 8216 $\AA$ line because the synthetic spectra do not
properly reproduce the observed stellar pseudo-continuum in this spectral
region. Regarding the Ca absorption lines, we checked the strongest Ca I lines
like those at 6122, 6162, 6439, 6463 and 6573 $\AA$ as well as the Ca II triplet
at longer wavelengths ($\sim$ 8500 $\AA$). The Ca I 6463 $\AA$ line turned out
to  be the best Ca abundance indicator. This is because the synthetic fits
around the  Ca I 6463 $\AA$ line (i.e. the stellar pseudo-continuum) are much
better than for the rest of Ca I lines; the stronger Ca I 6573 $\AA$ line also
displays saturation effects. As we have mentioned in the Introduction, the
isotopic Ca composition is theoretically expected to be affected by neutron
captures. The Ca isotope ratios cannot be measured from the atomic Ca absorption
lines (the atomic lines are intrinsically too broad; even at very
high-resolution). To comfort our Ca measurement from atomic lines, we
additionally explored the possibility of detecting the most intense CaH
\cite{shayesteh13,alavi18} and CaO \cite{yurchenko16} bandheads (around
$\sim$6850-6950 and 8650-8850 $\AA$, respectively) in the optical spectra of our
sample stars. Unfortunately, spectral synthesis show that no CaH and CaO
molecular lines are detectable in these spectral regions, which are completely
dominated by TiO. 

Thus, the Li and Ca abundances were determined from the Li I 6708 $\AA$ and Ca I
6463 $\AA$ lines, respectively, by using extended model atmospheres developed by
us \citep[see][for further information]{zamora14,perez-mesa17}. The atmospheric
parameters (T$_{eff}$ and log$g$), additional observational information
(variability period and OH expansion velocity) and the Li abundance derived
from hydrostatic models are listed in Table \ref{table_obs_param} for our sample
stars \citep[see][for more details]{garcia-hernandez07,garcia-hernandez13}.

\begin{table*}[]
\begin{center}
\caption{Atmosphere parameters and Li abundances (derived using hydrostatic models) and other selected observational information.}
\label{table_obs_param}
\renewcommand{\arraystretch}{1.25}
\footnotesize
\begin{tabular}{cccccccc}
\hline
\hline
IRAS name  & $T_{eff}$ (K) & log $g$ & v$_{exp}$(OH) (km s$^{-1}$) & Ref. & Period (days) & Ref. & log $\varepsilon(Li)_{static}$                            \\
\hline
01085$+$3022   &   3300   &   $-$0.5   &   13 & 1 &   560 & 1 &   2.4       \\
02095$-$2355   &   3300   &   $-$0.5   &   12{\textsuperscript{$\dagger$}} & ...  &   659 & 2  &   1.6      \\
%04404$-$7427{\textsuperscript{*}}   &   3000   &   $-$0.5   &   8 & 2 &   534 & 3  &   ...    \\
05027$-$2158   &   2800   &   $-$0.5   &   8 & 2 &   368 & 1 &   1.1         \\
05098$-$6422   &   3000   &   $-$0.5   &   6 & 3 &   394 & 3 &   $\leq-$1.0      \\
05151$+$6312   &   3000   &   $-$0.5   &   15 & 3  &   628 & 4  &   $<$   0.0      \\
05559$+$3825   &   2900   &   $-$0.5   &   12{\textsuperscript{$\dagger$}} & 6 &   590 & 5 &   0.6          \\
06300$+$6058   &   3000   &   $-$0.5   &   12 & 3 &   440 & 6  &   0.7          \\
%07222$-$2005{\textsuperscript{*}}   & 3000   &   $-$0.5   &   8 & 2  &   1200 & 7  &   ...       \\
07304$-$2032   &   2700   &   $-$0.5   &   7  & 4 &   509 & 1  &   0.9         \\
%09194$-$4518{\textsuperscript{**}}   &   3000   &   $-$0.5   &   11 & 2   &   ...   &   ... & ...      \\
09429$-$2148   &   3300   &   $-$0.5   &   12  & 3 &   650 & 1  &   2.2       \\
10261$-$5055   &   3000   &   $-$0.5   &   4  & 2 &   317 & 1  &   $\leq-$1.0      \\
11081$-$4203   &   3000   &   $-$0.5   &   8{\textsuperscript{$\dagger$}} & 2 &   332 & 8  &   1.3         \\
%11525$-$5027   &   3300   &   $-$0.5   &   ...  & ... & ... &  ...   &   0.9              \\
%12377$-$6102{\textsuperscript{*}}   &   ...   &   $-$0.5   &   20 & 2 &   ... & ...  &   ...      \\
14266$-$4211   &   2900   &   $-$0.5   &   9  & 2 &   389 & 8  &   $\leq$ 0.0         \\
14337$-$6215   &   3300   &   $-$0.5   &   20  & 5 &   ... & ...  &   2.4{\textsuperscript{1}}         \\
15193$+$3132   &   2800   &   $-$0.5   &   3  & 3 &   360  & 1 &   $\leq$ 0.0         \\
15211$-$4254   &   3300   &   $-$0.5   &   11 & 2 &   ... & ...  &   2.3              \\
15255$+$1944   &   2900   &   $-$0.5   &   7  & 3 &   425 & 1  &   1.0         \\
15576$-$1212   &   3000   &   $-$0.5   &   10 & 3  &   415 & 1  &   1.1           \\
16030$-$5156   &   3000   &   $-$0.5   &   10{\textsuperscript{$\dagger$}} & 2 &   579 & 9 &   1.5          \\
16037$+$4218   &   2900   &   $-$0.5   &   4  & 2 &   360 & 10 &   $\leq-$1.0         \\
16260$+$3454   &   3300   &   $-$0.5   &   12 & 3  &   475 & 1 &   2.7         \\
17034$-$1024   &   3300   &   $-$0.5   &   8{\textsuperscript{$\dagger$}} & 2 &   346 & 1 &   $\leq$ 0.0        \\
%18025$-$2113{\textsuperscript{1}}   &   ...   &   $-$0.5   &   11 & 2 &   828 & 8 &   ...      \\
%18057$-$2616{\textsuperscript{2}}   &   3000   &   $-$0.5   &   19 & 2 &   720   & 8 & ...           \\
18413$+$1354   &   3300   &   $-$0.5   &   15 & 6 &   590 & 5 &   1.8                      \\
18429$-$1721   &   3000   &   $-$0.5   &   7 & 2 &   481 & 8 &   1.2               \\
%19059$-$2219{\textsuperscript{**}}   &   3000   &   $-$0.5   &   13 & 3 &   510 & 1 &   ...           \\
19129$+$2803 & 3300 & $-$0.5 & 11{\textsuperscript{$\dagger$}} & 2 & 420 & 10 &   3.1{\textsuperscript{1}}  \\
19361$-$1658   &   3000   &   $-$0.5   &   8  & 2 &   ... & ...  &   1.9             \\
%19426$+$4342{\textsuperscript{**}}   &   3000   &   $-$0.5   &   9 & 2 &   ... & ...  &   ...             \\
20052$+$0554   &   3300   &   $-$0.5   &   16 & 7  &   450 & 5 &   2.6         \\
%20077$-$0625{\textsuperscript{**}}   &   3000   &    $-$0.5   &   12 & 3 &   680 & 1 &   ...          \\
20343$-$3020   &   3000   &   $-$0.5   &   8 & 2 &   349 & 1 &   $\leq-$1.0               \\
RU Cyg   &   3000   &   $-$0.5   &   12{\textsuperscript{$\dagger$}} & ... &   442 & 11  &   2.0        \\
SV Cas   &   3000   &   $-$0.5   &   12{\textsuperscript{$\dagger$}} & ... &   456 & 11 &   3.5              \\
R Cen    &   3000   &   $-$0.5   &   5{\textsuperscript{$\dagger$}} & ... &   251 & 11 &   4.3    \\
\hline
\end{tabular}
\end{center}
%\textsuperscript{*}\footnotesize{The S/N at Li I 6708 \AA~ is very low to derive any Li abundance estimate.}\\ 
%\textsuperscript{**}\footnotesize{Optical counterpart too red.} \\
%\textsuperscript{1}\footnotesize{Possible double lined spectroscopic binary.} \\
%\textsuperscript{2}\footnotesize{The Li I line has a P cygni-type profile.} \\
\textsuperscript{$\dagger$}\footnotesize{These stars only display a single peak in the 1612 MHz OH maser. Thus, in the abundance analysis, we adopted the average OH expansion velocities from the velocity values displayed by the other sample stars with similar variability periods.} \\
\textsuperscript{1}\footnotesize{The line is resolved in two components (circumstellar and photospheric) and the abundance obtained corresponds to the photospheric one.} \\
\footnotesize{References for the OH expansion velocities: \\
1) \cite{chengalur93}; 2) \cite{telintelhekkert91}; 3) \cite{telintelhekkert89}; 4) \cite{sivagnanam89}; 5) \cite{sevenster97}; 6) \cite{slootmaker85}; 7) \cite{lewis94}.} \\
\footnotesize{References for the periods: \\
1) Combined General Catalogue of Variable Stars (GCVS), \cite{kholopov98}; 2) \cite{richards12}; 3) \cite{whitelock94}; 4) \cite{wozniak04}; 5) \cite{jones90}; 6) \cite{lockwood85}; 7) \cite{groenewegen98}; 8) AAVSO International Variable Star Index, \cite{watson06}; 9) General Catalogue of Variable Stars (GCVS), \cite{samus17}; 10) \cite{jimenezesteban06}, 11) \cite{garcia-hernandez13}.} \\
%\textsuperscript{$\dagger$}\footnotesize{These stars only display a single peak in the 1612 MHz OH maser, so we adopt an average velocity in the abundace analysis exploring OH expansion velocities of other stars with similar variability periods.} \\
\end{table*}

%
%______________________________________________________________

\section{Pseudo-dynamical models}

In the chemical abundance analysis, we have followed the previous work by
\cite{perez-mesa17}. In short, we have used the v12.2 modified version of the 
spectral synthesis code \textit{Turbospectrum} \citep{alvarez98,plez12},  in
which the presence of a circumstellar gas envelope and a radial wind  are
considered. In the analysis of the stars in our sample, we have assumed the
atmosphere parameters (e.g. T$_{eff}$, log$g$, C/O, [Fe/H], macroturbulence) 
from \cite{garcia-hernandez07,garcia-hernandez13} and the solar abundances from
\cite{grevesse07}. Hydrodynamical wind models for AGB stars have been 
developed through the years \citep[see the review by][and references therein]{hofner18}.
Recent pulsation-enhanced dust-driven outflow type models include time-dependent 
gas dynamics and dust formation, with polychromatic radiative transfer 
\citep[e.g.][]{eriksson14, hofner16}. Their predictive power for a particular star 
is however limited by the use of free parameters, e.g. the piston velocity and 
amplitude driving the pulsations. We therefore chose to use generic empirical models, 
based on observational determinations of the velocity-law and simple physical 
hypotheses. The pseudo-dynamical models were constructed from the
original MARCS hydrostatic model structure and the atmosphere radius was
extended by a wind out to $\sim$5 stellar radii and a radial velocity field. We
have computed the stellar wind following mass conservation, radiative
thermal equilibrium and a classical $\beta$-velocity law 
\citep[see][for more details]{zamora14,perez-mesa17}.

By adopting the atmospheric parameters from \cite{garcia-hernandez06,
garcia-hernandez07,garcia-hernandez13}, we genereted a mini-grid of 
synthetic spectra for each sample star. Some parameters are fixed: 
stellar mass M = 2  M$_{\odot}$\footnote{The stellar mass was in all 
cases selected to be 2 M$_{\odot}$ because the temperature and pressure 
structure of the model atmosphere is pratically identical for a 1 
M$_{\odot}$ and 10 M$_{\odot}$ model atmosphere and the output synthetic 
spectra are not sensitive to the mass of the star \citep[see Fig. 1 in][]{plez90}.}, gravity log $g$ = $-$0.5, 
microturbulent velocity $\xi$ = 3  km/s, metallicity [Fe/H] = 0.0 and C/O =
0.5 dex \citep[see][for more details]{garcia-hernandez07}. However, for the
mass-loss rate $\dot{M}$ and $\beta$ parameters, we use values between $\dot{M}
\sim 10^{-9}-10^{-6}  M_{\odot}yr^{-1}$ \footnote{Massive AGB stars with
mass-loss rates higher than 10$^{-6}$ M$_{\odot}yr^{-1}$ are completely obscured
in the optical, while our sample stars, still visible in the optical, should
have lower mass-loss rates \citep[see][for more details]{zamora14,perez-mesa17}.} in steps of $0.5\times10^{-1}$
$M_{\odot}yr^{-1}$ and $\beta \sim 0.2-1.6$ in steps of 0.2, respectively.
Finally, for the Li and Ca abundances we used values between log
$\varepsilon(Li)$ $\sim$ 0.0 to $+$2.8  dex and log $\varepsilon(Ca)$ $\sim$
$+$5.0 to $+$7.0 dex, in steps of 0.1 dex. The parameters of the synthetic
spectra that best fit the 6708 \AA~Li I and  the 6463 \AA~Ca I profiles and
their adjacent pseudocontinua are listed in Table \ref{table_abundances}.

For the subsample of stars (15) from \cite{perez-mesa17} we have used the
stellar and wind parameters obtained from the Rb I 7800 \AA~spectral region fits
for consistency and because the synthetic spectra are much less sensitive to
variations of the model parameters in the Li I 6708 \AA~and Ca I 6463 \AA~spectral
regions (see Section 3.1 below). 

%\textbf{The mass-loss rates from CO available in the literaure  strongly depend
%of the authors and the $\dot{M}$ relation used for the sources,  and also, the
%mass-loss rate could not be constant \citep[e.g.][]{delfosse97}.  The
%mass-loss rates from the literaure are sistematically one order magnitude 
%larger than the obtained in \cite{perez-mesa17}, converting these massive
%Rb-rich  AGB stars into Rb-poor ones, at odds with the more recent AGB
%nucleosynthesis  theoretical models: \cite{vanraai12}, \cite{karakas12},
%\cite{karakaslugaro16}  (Monash), \cite{ritter18} (NuGrid/MESA),
%\cite{cristallo15} (FRUITY).}

\subsection{Sensitivity of the synthetic spectra to variations of the model parameters}

\begin{figure*}
   \centering
   \includegraphics[width=9.15cm,height=6.7cm,angle=0]{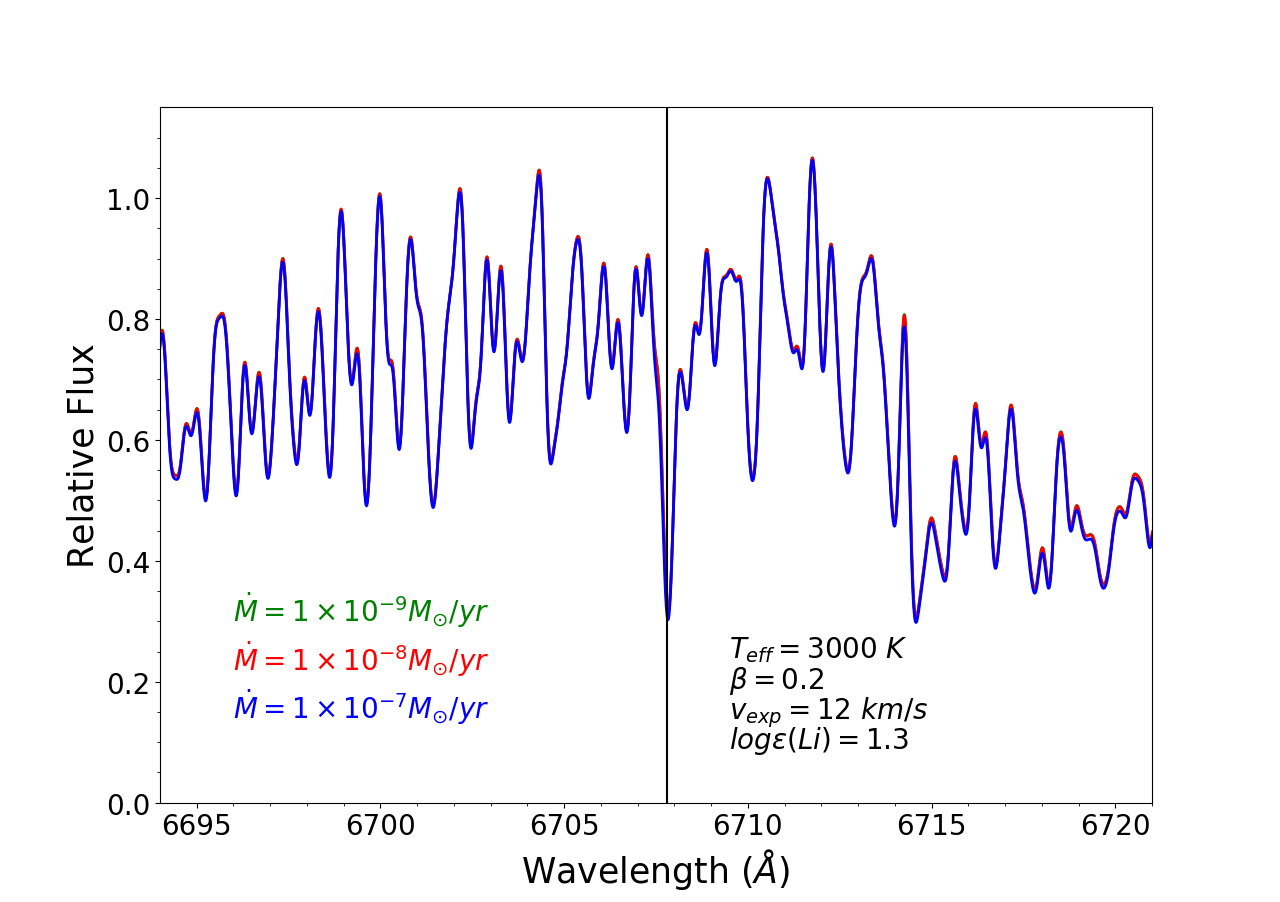}
   \includegraphics[width=9.15cm,height=6.7cm,angle=0]{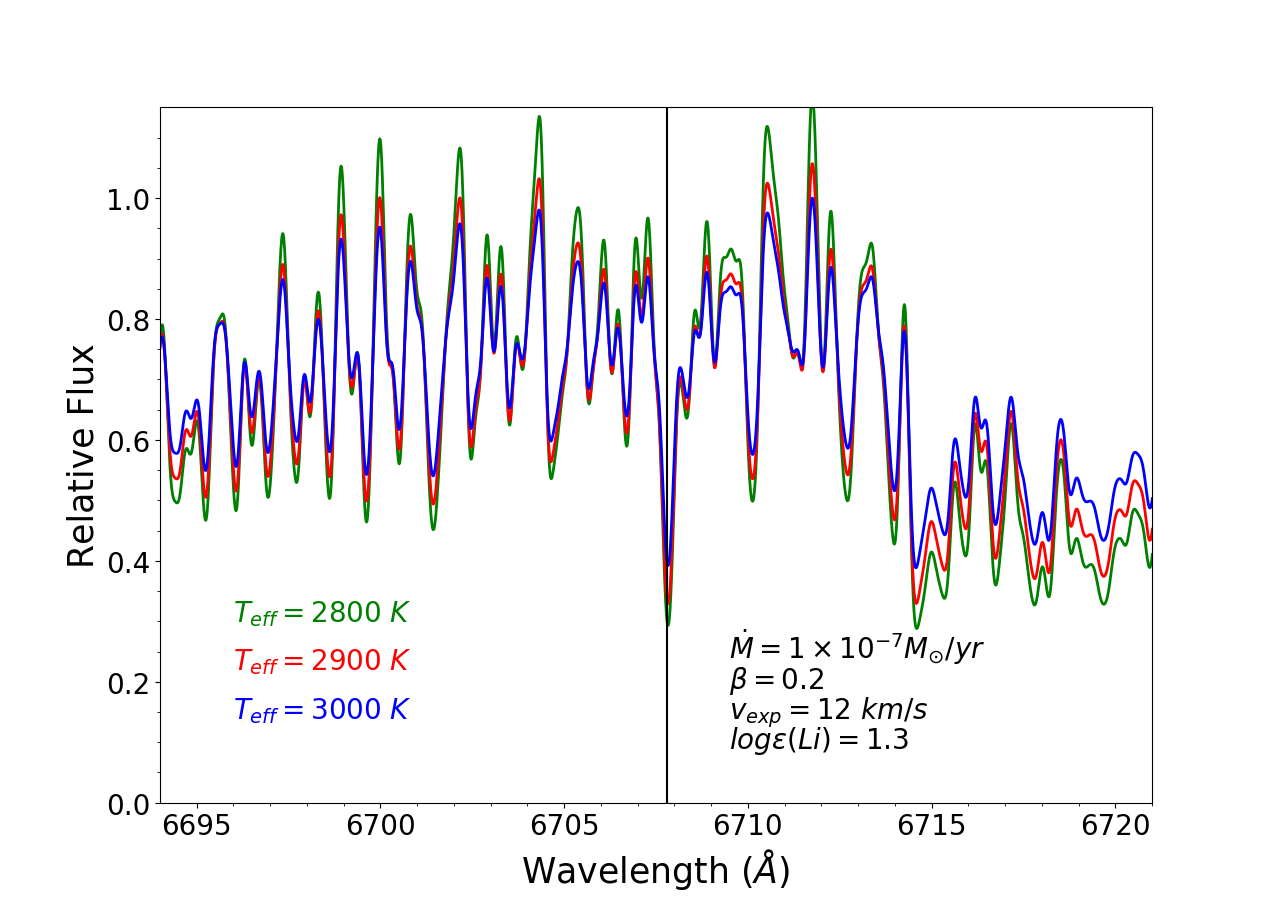}
   \includegraphics[width=9.15cm,height=6.7cm,angle=0]{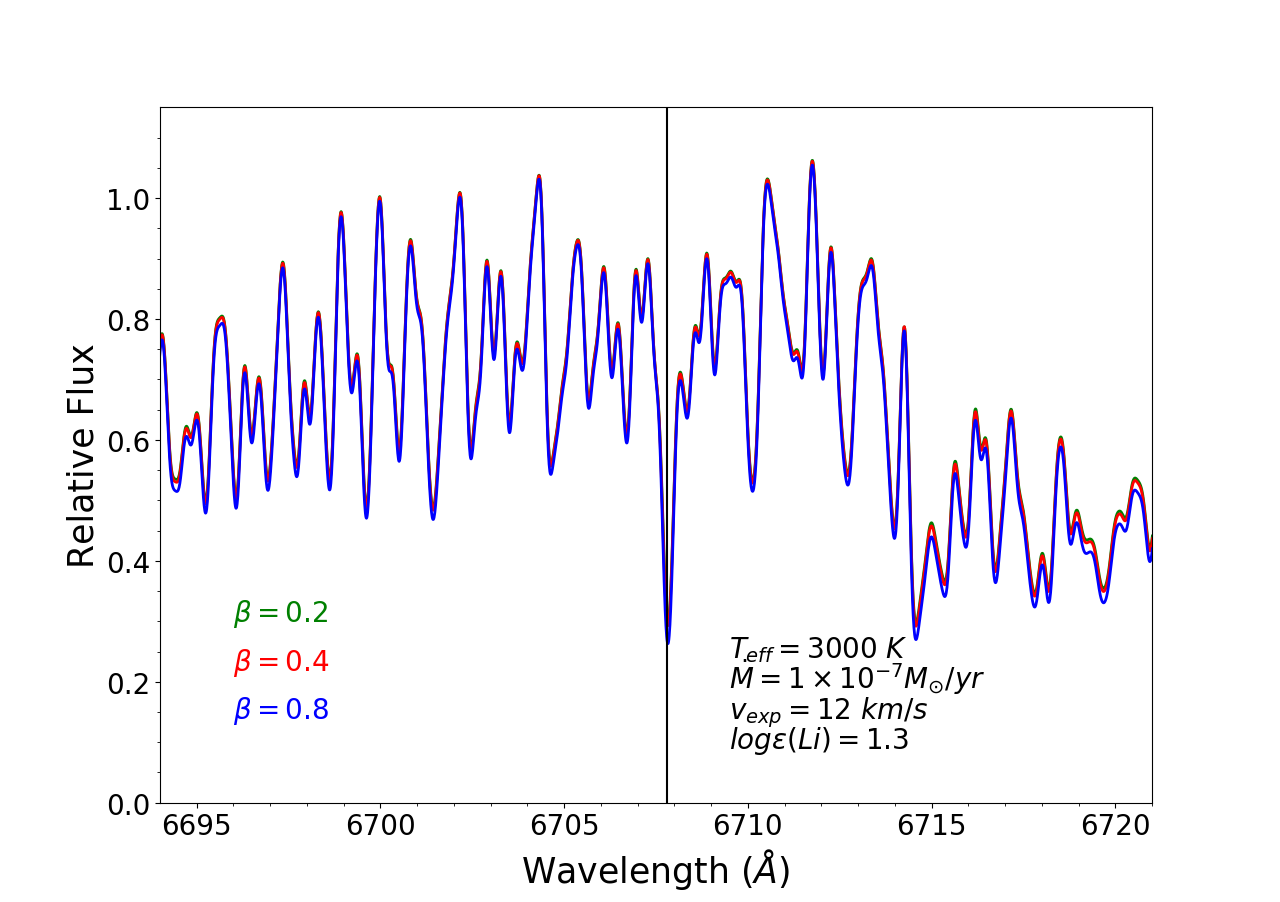}
   \includegraphics[width=9.15cm,height=6.7cm,angle=0]{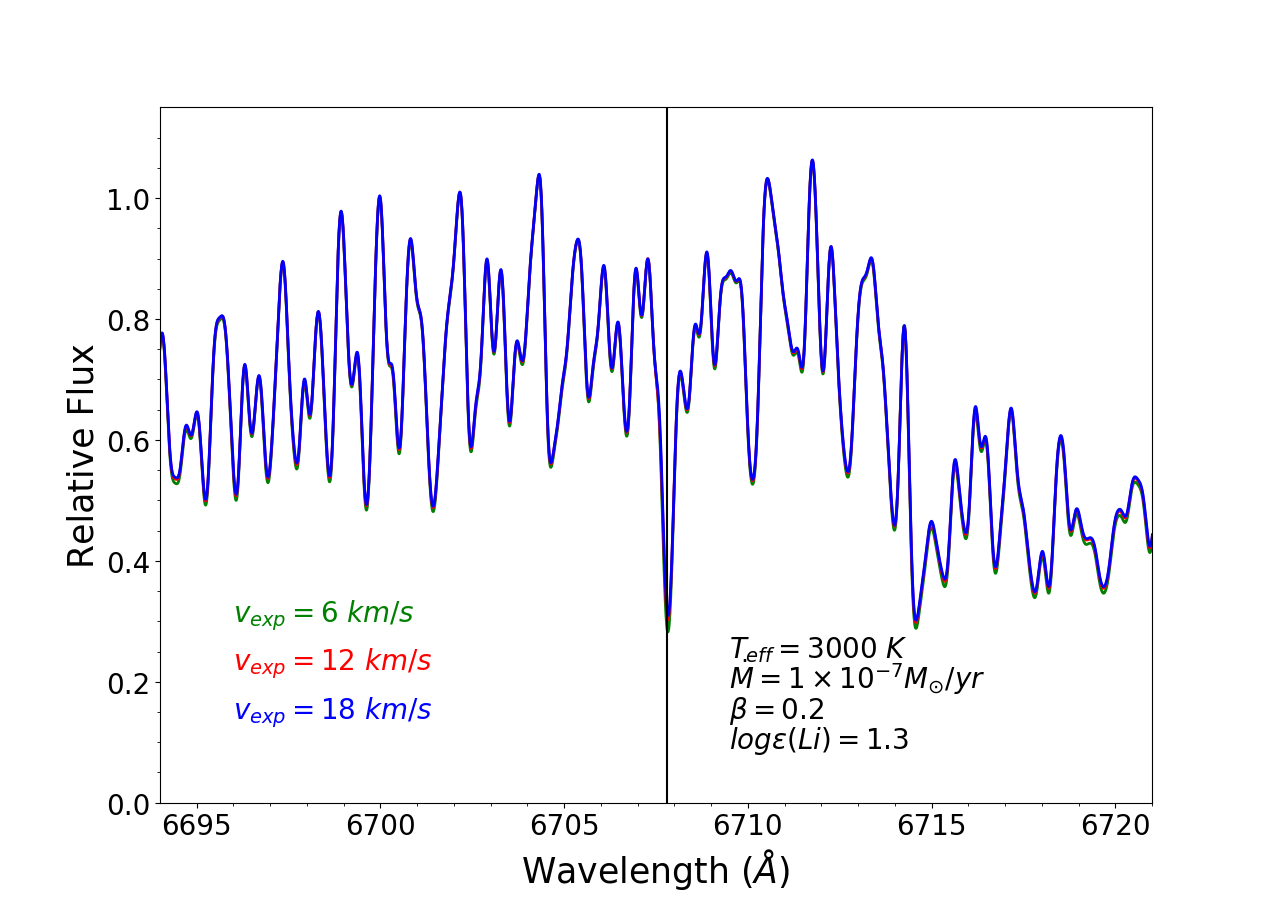}
      \caption{Illustrative examples of synthetic spectra for different stellar
(T$_{eff}$) and wind ($\dot{M}$, $\beta$, and $v_{exp}$(OH)) parameters in the
spectral region around the 6708 $\AA$ Li I line. The gravity log $g$ = $-$0.5 
is the same in all the spectra. The black vertical line indicates the position of the 6708 
$\AA$ Li I line.}
         \label{comparisons_Li}
\end{figure*}

\begin{figure*}
   \centering
   \includegraphics[width=9.15cm,height=6.7cm,angle=0]{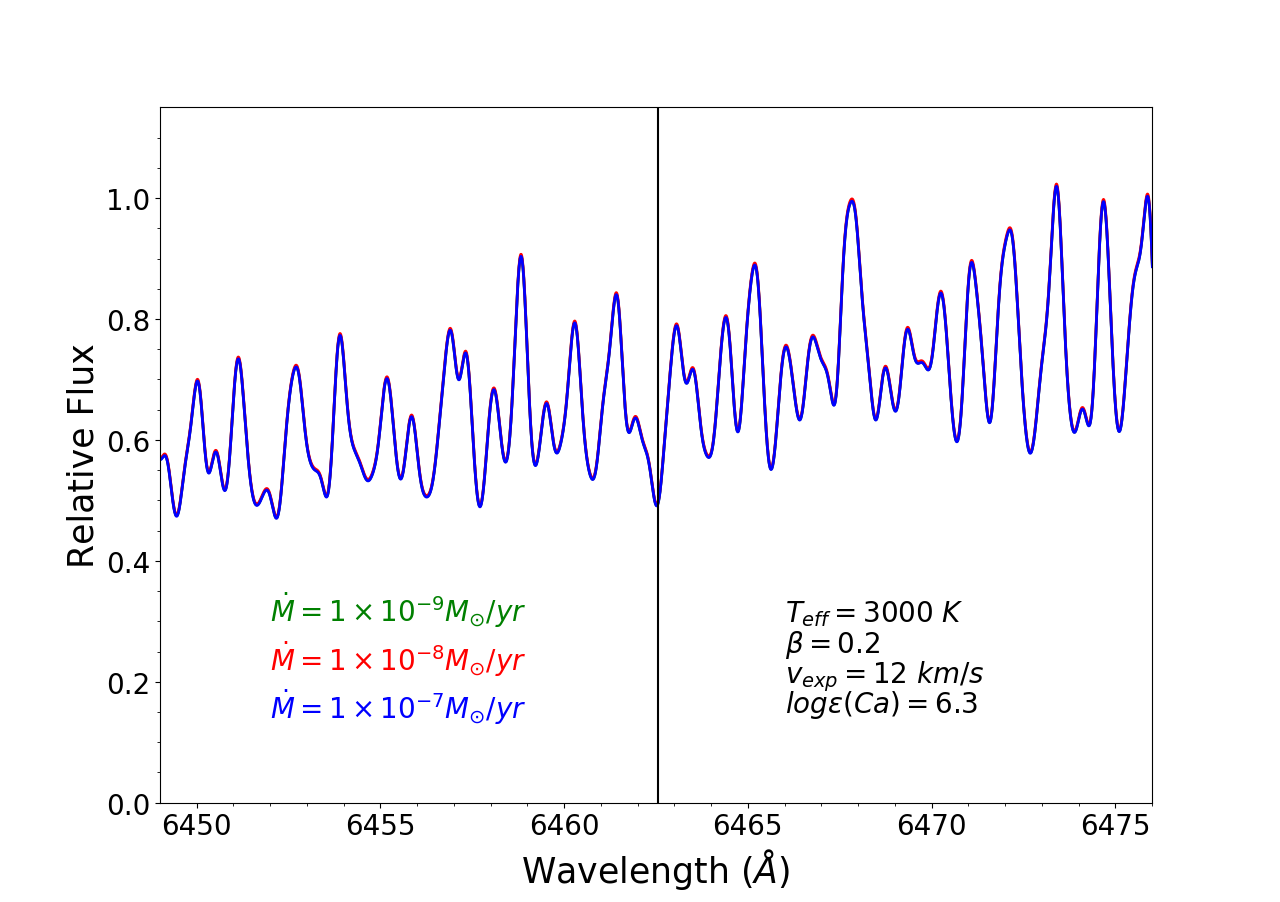}
   \includegraphics[width=9.15cm,height=6.7cm,angle=0]{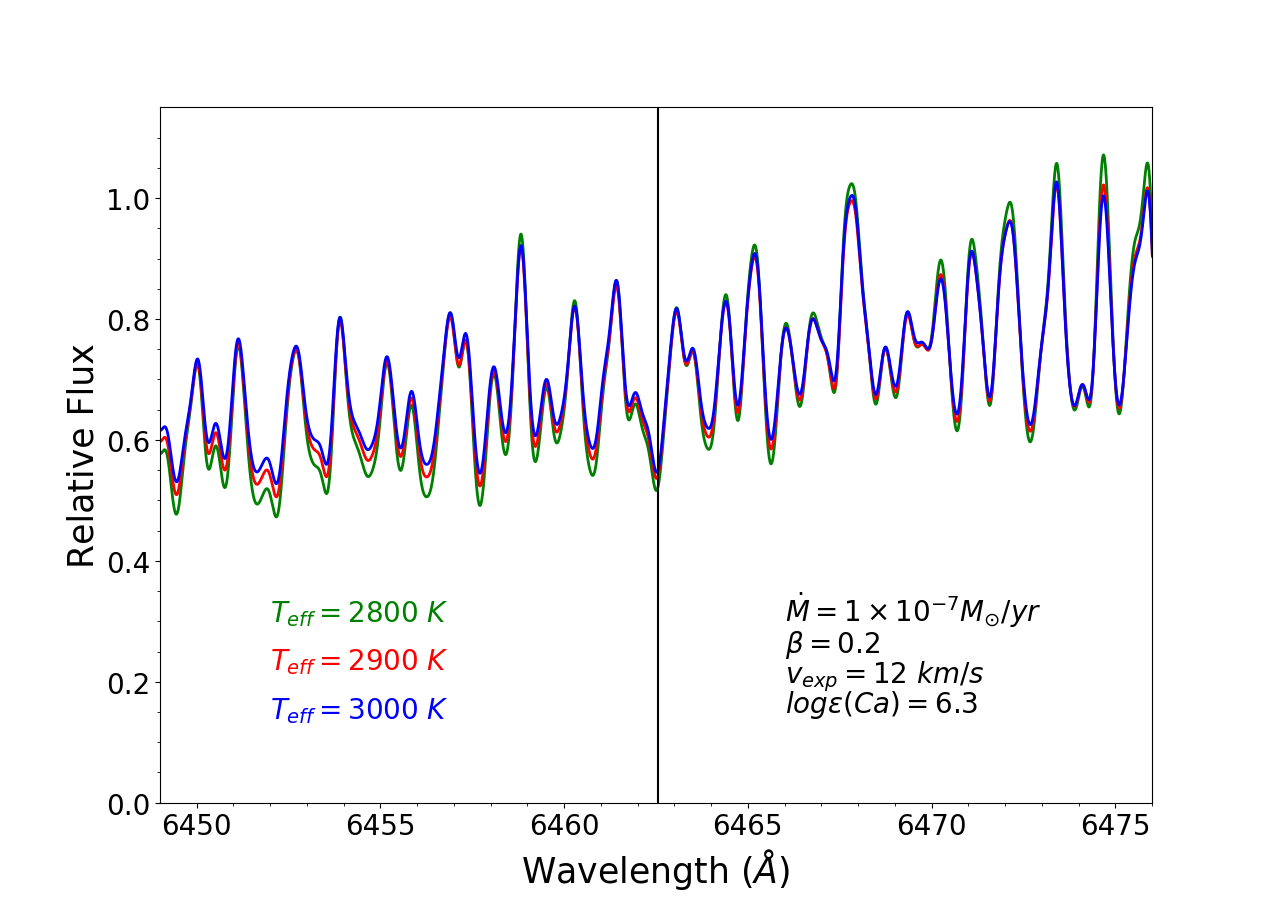}
   \includegraphics[width=9.15cm,height=6.7cm,angle=0]{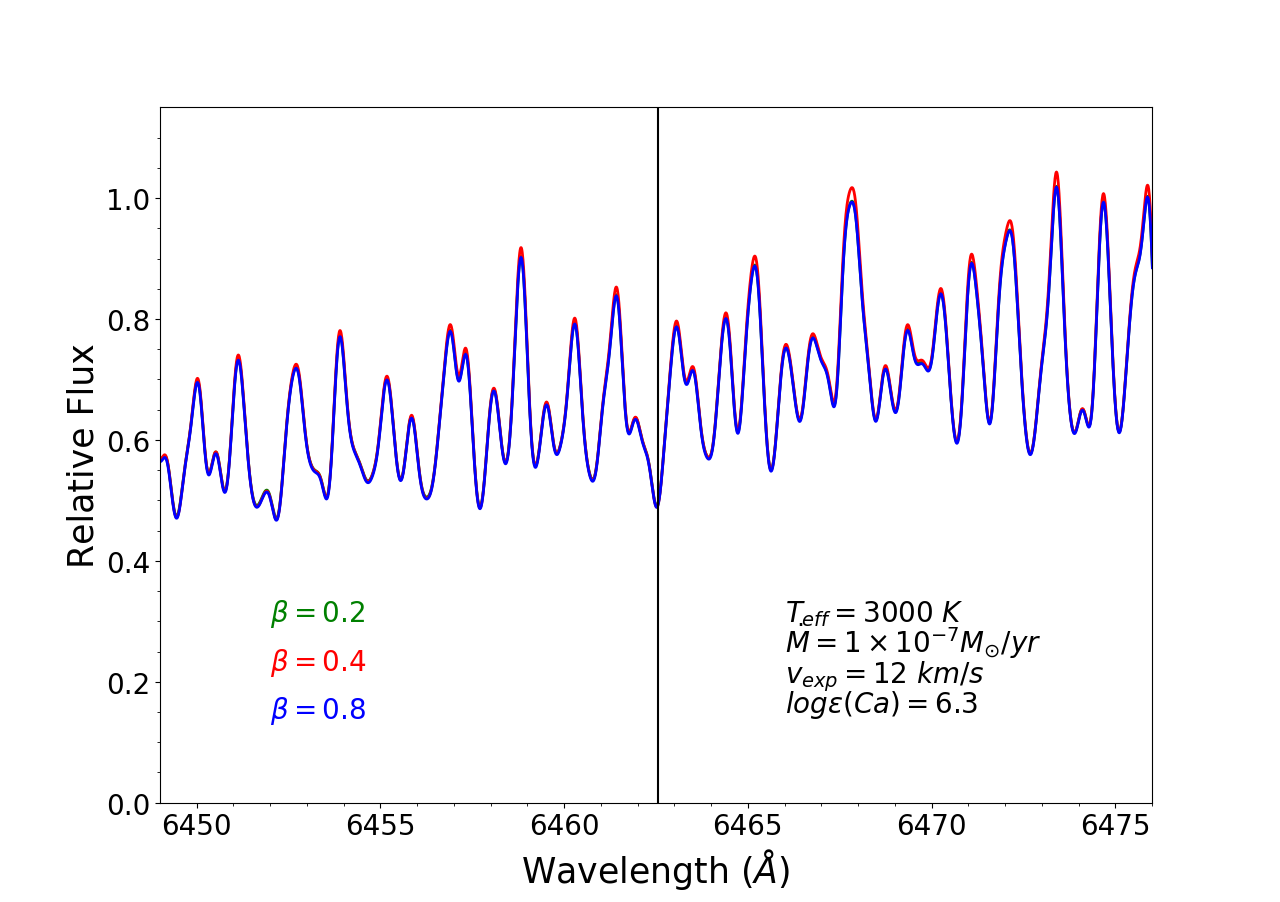}
   \includegraphics[width=9.15cm,height=6.7cm,angle=0]{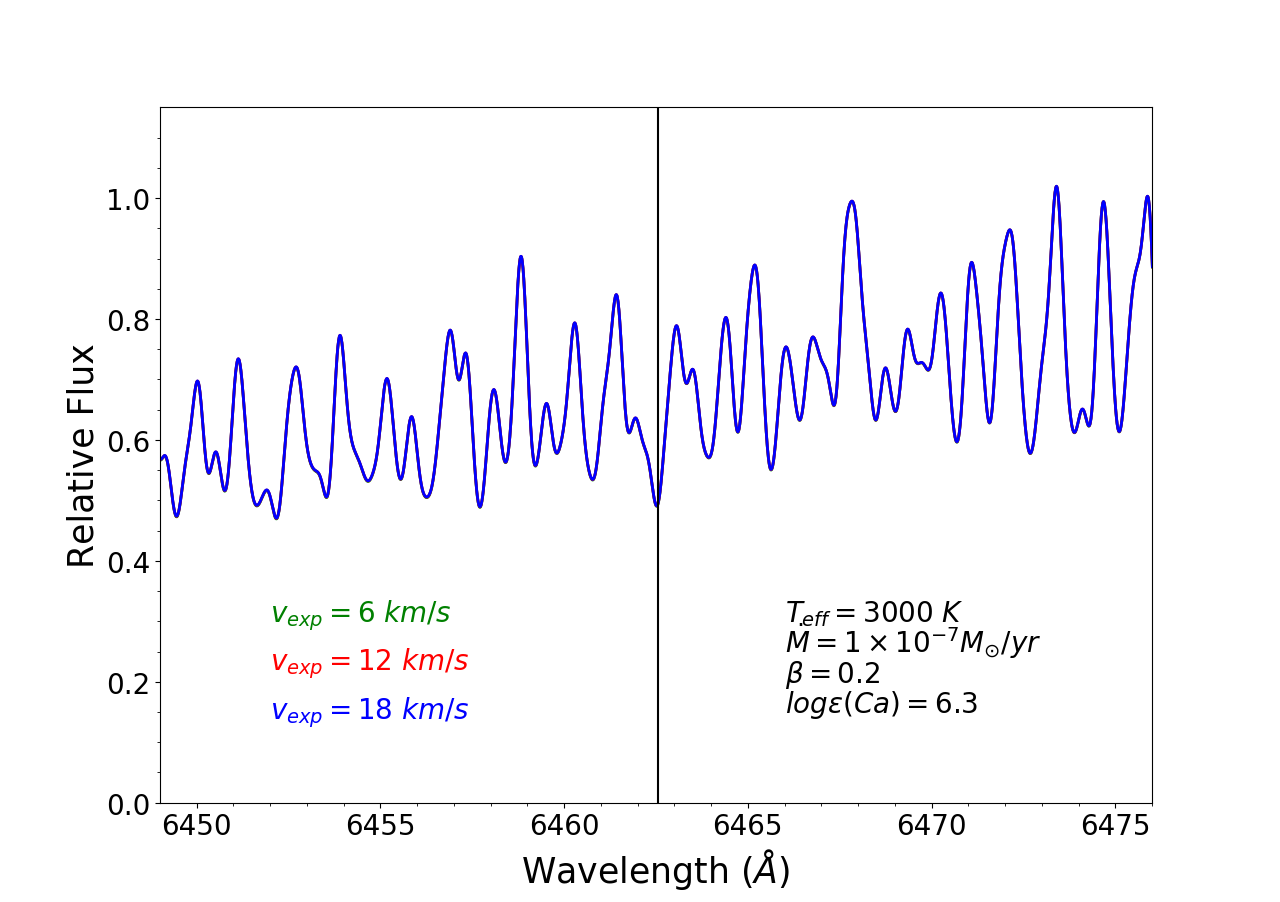}
      \caption{Illustrative examples of synthetic spectra for different stellar
(T$_{eff}$) and wind ($\dot{M}$, $\beta$, and $v_{exp}$(OH)) parameters in the
spectral region around the 6462.6 $\AA$ Ca I line. The gravity log $g$ = $-$0.5 
is the same in all the spectra. The black vertical line indicates the position of the 6462.6 
$\AA$ Li I line.}
         \label{comparisons_Ca}
\end{figure*}

We have analyzed the influence of variations in the stellar (T$_{eff}$) and wind
($\dot{M}$, $\beta$ and $v_{exp}$(OH)) parameters in the output synthetic
spectra. Figures \ref{comparisons_Li} and \ref{comparisons_Ca} show,
respectively, synthetic spectra in the spectral regions around the 6708 $\AA$ Li
I and 6463 $\AA$ Ca I lines for different stellar and wind parameters.

The Li I profile is not very sensitive to the wind parameters ($\dot{M}$,
$\beta$  and $v_{exp}$(OH)). The Li I line is only slightly stronger with
increasing $\dot{M}$ (Figure \ref{comparisons_Li}; top-left panel) and $\beta$
(Figure \ref{comparisons_Li};  bottom-left panel), while it is slightly
weaker with increasing $v_{exp}$(OH) (Figure \ref{comparisons_Li}; bottom-right
panel). In addition, the Li I absorption line is stronger with decreasing 
T$_{eff}$ (Figure \ref{comparisons_Li}; top-right panel), with the
pseudo-continuum (e.g. the TiO molecular bands) being also affected, as
expected \citep[see e.g.][]{garcia-hernandez07}. In the Ca case, the
sensitivity of the synthetic spectra to variations in $\dot{M}$, $\beta$ and
$v_{exp}$(OH) (Figure \ref{comparisons_Ca}; top-left,  bottom-left and
bottom-right panel, respectively) is even smaller than the Li case. The Ca I
6463 \AA~spectral region displays TiO molecular bands weaker than the Li I
6708 \AA~spectral region and, consequently, the Ca I line and the
pseudo-continuum are less affected by T$_{eff}$ variations (Figure
\ref{comparisons_Ca}; top-right  panel) than in the Li I region.

%
%______________________________________________________________

\section{Abundance results}

The parameters of the best fits of \cite{garcia-hernandez06, 
garcia-hernandez07, garcia-hernandez13} to the observations and the hydrostatic Li 
abundances are listed in Table \ref{table_obs_param}. In these fits, the static models
have used the solar abundances from \cite{grevesse98} for computing the Li abundances,
while the new pseudo-dynamical models and the hydrostatic values shown in Table 
\ref{table_abundances} use the more recent solar abundances from \cite{grevesse07}.
The Li hydrostatic abundances obtained by using \cite{grevesse98} and 
\cite{grevesse07} are practically the same. 

\begin{figure*}
\centering
\includegraphics[width=9.1cm,height=6.7cm,angle=0]{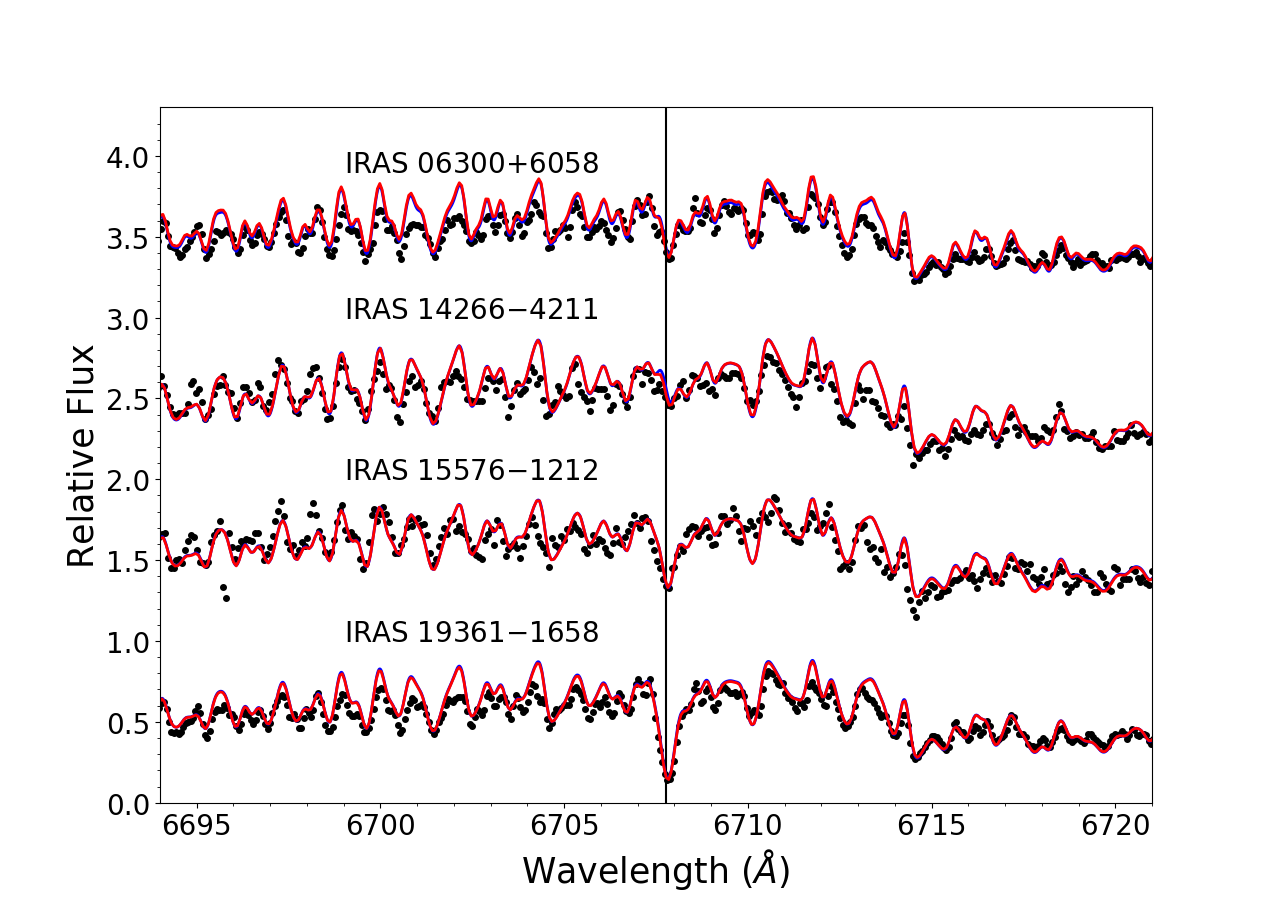}
\includegraphics[width=9.1cm,height=6.7cm,angle=0]{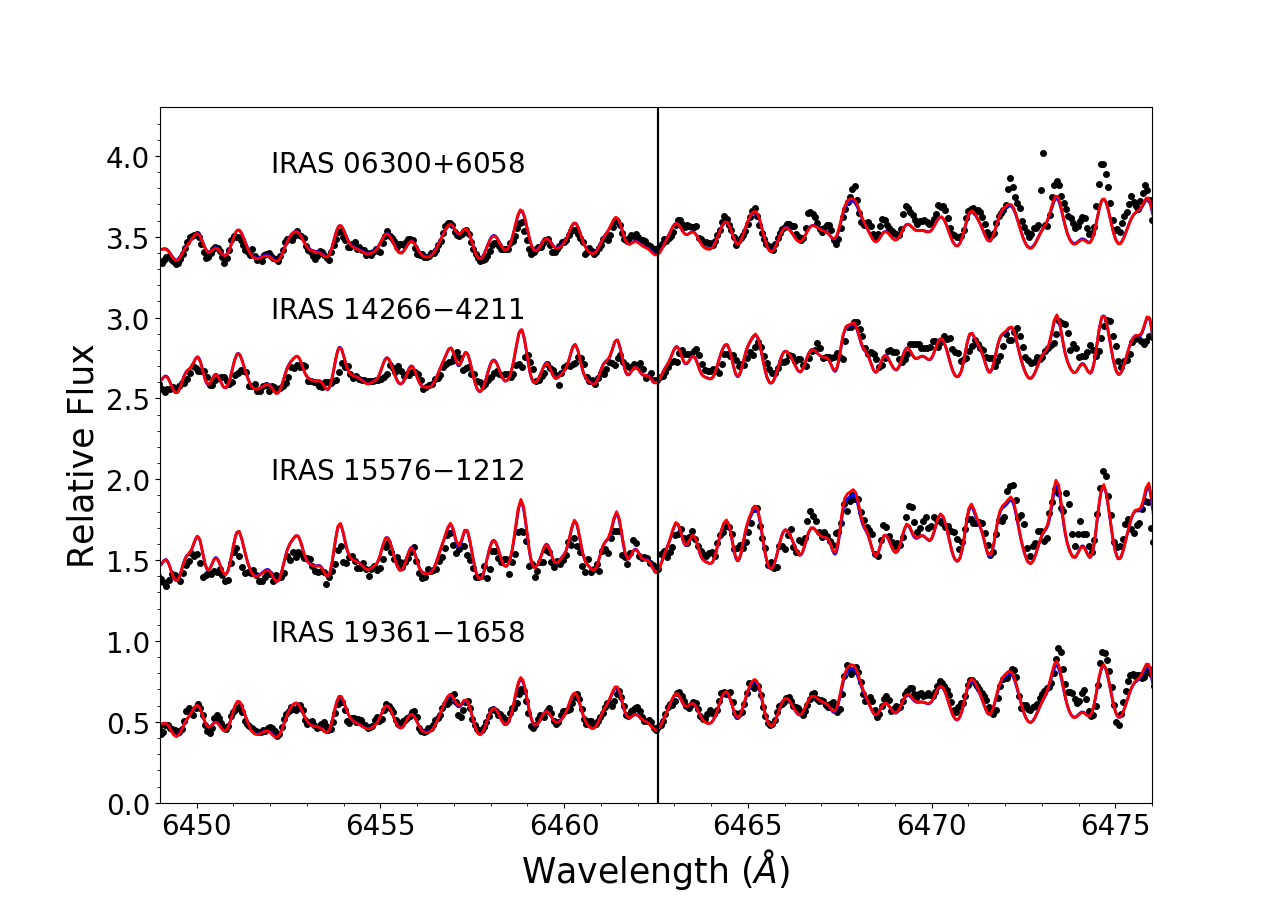} 
\caption{The Li I 6708 $\AA$ (\textit{left panel}) and Ca I 6463 $\AA$
(\textit{right panel}) spectral regions in four massive Galactic AGB stars. The
hydrostatic models (blue lines) and the pseudo-dynamical models (red lines) 
that best fit the observations (black dots) are shown. The location of the Li 
I and the Ca I lines are indicated by black vertical lines.}
\label{Li_Ca_line}
\end{figure*}

\begin{figure*}
\centering
\includegraphics[width=9.1cm,height=6.7cm,angle=0]{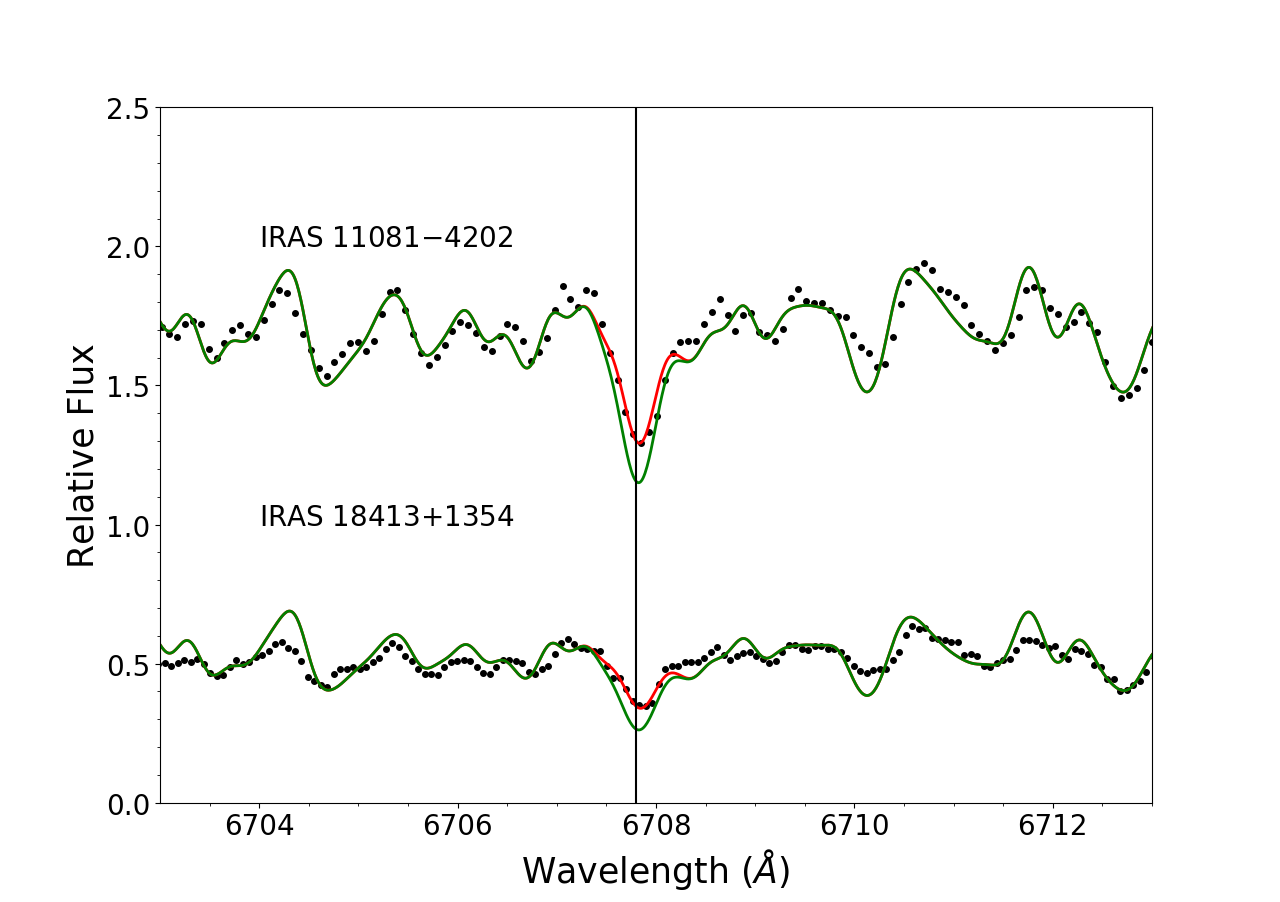}
\includegraphics[width=9.1cm,height=6.7cm,angle=0]{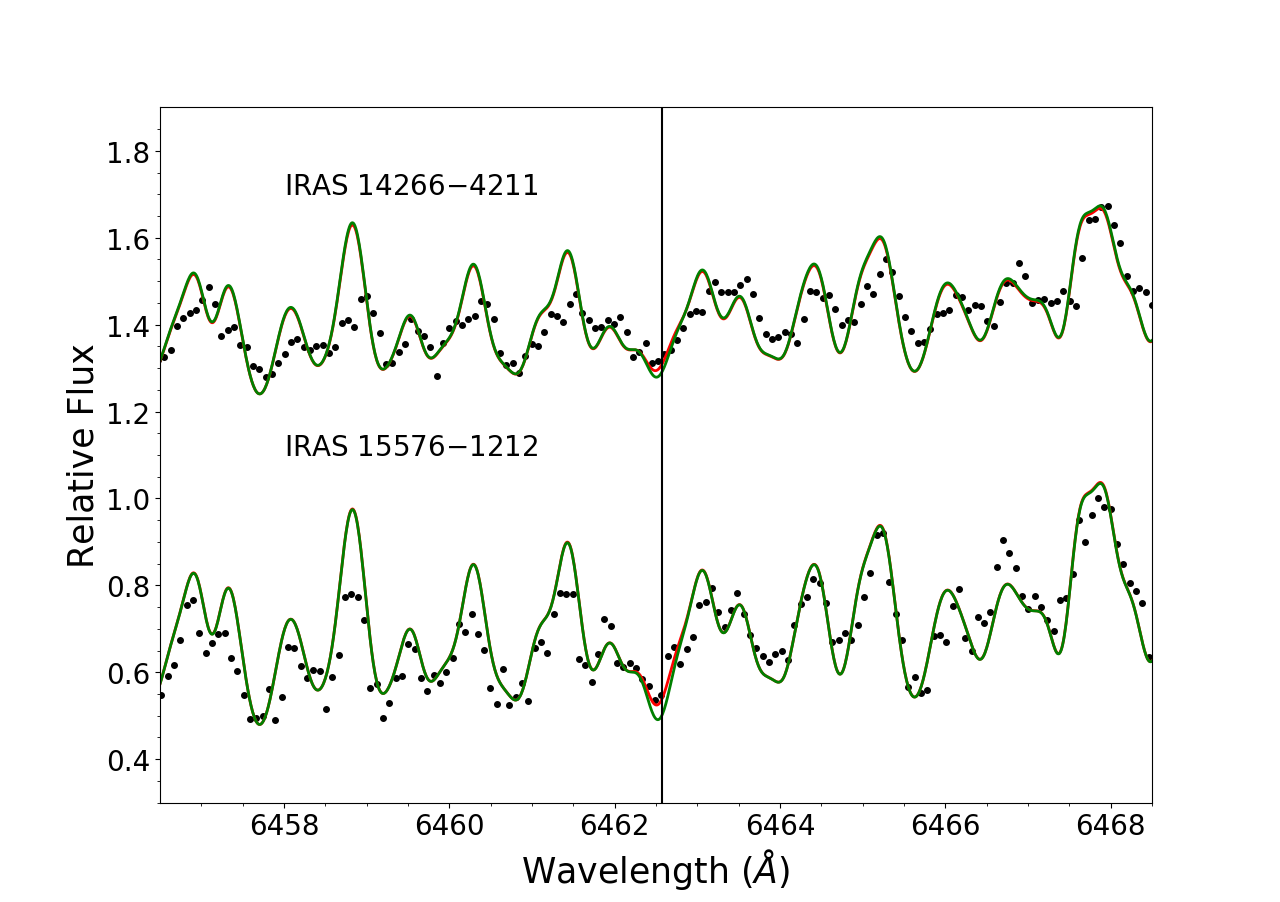}
\caption{A zoom in the Li I 6708 $\AA$ (\textit{left panel}) and Ca I 6463 $\AA$
(\textit{right panel}) spectral regions. The pseudo-dynamic  models (red lines)
that best fits the observations (black dots) are shown together with the
corresponding synthetic spectra for Li and Ca abundances 0.5 dex higher (green
lines).}
\label{Li_Ca_zoom}
\end{figure*}

In Figure \ref{Li_Ca_line} we display the hydrostatic and pseudo-dynamical fits
in the 6708 $\AA$ Li I and 6463 $\AA$ Ca I regions in four sample stars. The
pseudo-dynamical models are similar to the hydrostatic ones, and  reproduce
properly the Li and Ca regions. The Li and Ca line profiles are not  strongly
affected by the presence of a circumstellar envelope and a radial wind. The rest
of spectral fits are shown in Figure \ref{Li_sample} and \ref{Ca_sample} in 
Appendix A. In addition, Figure  \ref{Li_Ca_zoom}
displays in more detail the Li and Ca regions for some sample stars, showing
that the Ca I line is less sensitive (but still useful) to abundance variations
than the Li one. In seven stars (IRAS 02095$-$2355, IRAS 09429$-$2148, IRAS
15211$-$4254, IRAS 16260$+$3454, IRAS 17034$-$1024, IRAS  18413$+$1354 and IRAS
19129$+$2803) the best spectral fit is different in the Li and Ca regions. In
all cases, around the 6708 $\AA$ Li I line the best spectral fits give
$T_{eff}$  = 3300 K, while around the 6463 $\AA$ Ca I line the best spectral
fits provide cooler $T_{eff}$ of 3000 K. A similar finding, this time when
comparing the Li I 6708$\AA$ and Rb I 7800$\AA$ regions was previously found by
\cite{garcia-hernandez06,garcia-hernandez07}.

In addition, for some sample stars in which the OH expansion velocity is unknown 
(IRAS 02095$-$2355, IRAS 05559$+$3825, IRAS 11081$-$4203, IRAS 16030$-$5156, IRAS
17034$-$1024, IRAS 19129$+$2803, RU Cyg, SV Cas and R Cen), we explored the OH expansion 
velocity range displayed by other sample stars with similar variability periods. Due to 
the similar spectral fits that are obtained for slightly different OH expansion velocities, 
for these stars we thus adopted the average OH expansion velocities from the values displayed 
by the sample stars with similar periods (see Table \ref{table_obs_param}).

\begin{table*}[]
\begin{center}
\caption{Atmosphere parameters and best-fit Li and Ca pseudo-dynamical abundances for the sample of massive AGB stars.}
\label{table_abundances}
\renewcommand{\arraystretch}{1.25}
\footnotesize
\begin{tabular}{ccccccccccc}
\hline
\hline
IRAS name  & $T_{eff}$ (K) & log $g$ & $\beta$ & $\dot{M}$ (M$_{\odot}$ yr$^{-1}$) & v$_{exp}$(OH) (km s$^{-1}$) 
& log $\varepsilon(Li)_{static}$ & log $\varepsilon(Li)_{dyn}$ & log $\varepsilon(Ca)_{static}$ & log $\varepsilon(Ca)_{dyn}$  \\
\hline
01085$+$3022{\textsuperscript{*}} & 3300 & $-$0.5 & 0.2 & 1.0$\times$10$^{-7}$ & 13 & 2.4 & 2.2 & ... & ...      \\
02095$-$2355 & 3300{\textsuperscript{**}} & $-$0.5 & 0.8 & 1.0$\times$10$^{-9}$ & 12{\textsuperscript{$\dagger$}}&1.6 & 1.6 & 6.3 & 6.3 \\
%04404$-$7427{\textsuperscript{*}} & 3000 & $-$0.5 & ... & ... & 8 & ... & ... & ... & ...   \\
05027$-$2158 & 2800 & $-$0.5 & 0.4 & 1.0$\times$10$^{-7}$ & 8 & 1.1 & 1.1 & 5.8 & 5.8    \\
05098$-$6422 & 3000 & $-$0.5 & 1.4 & 1.0$\times$10$^{-8}$ & 6 & $\leq-$1.0 & $\leq$ 0.0 & 6.1 & 6.1 \\
05151$+$6312 & 3000 & $-$0.5 & 1.0 & 1.0$\times$10$^{-8}$ & 15 & $\leq$ 0.0 & $\leq$ 0.2 & 6.0 & 6.0  \\
05559$+$3825 & 2900 & $-$0.5 & 1.6 & 1.0$\times$10$^{-7}$ &12{\textsuperscript{$\dagger$}}& 0.6 & 0.5 & 5.8 & 5.8\\
06300$+$6058 & 3000 & $-$0.5 & 0.2 & 1.0$\times$10$^{-7}$ & 12 & 0.7 & 0.8 & $\leq$ 5.8 & $\leq$ 5.8  \\
%07222$-$2005{\textsuperscript{*}} & 3000 & $-$0.5 & ... & ... & 8 & ... & ... & ... & ... \\
07304$-$2032 & 2700 & $-$0.5 & 0.4 & 1.0$\times$10$^{-7}$ & 7 & 0.9 & 1.0 & $\leq$ 5.8 & $\leq$ 5.8  \\
%09194$-$4518{\textsuperscript{**}} & 3000 & $-$0.5 & ... & ... & 11 & ... & ... & ... & ...  \\
09429$-$2148 & 3300{\textsuperscript{**}} & $-$0.5 & 1.6 & 1.0$\times$10$^{-8}$ & 12 & 2.2 & 2.2 & 5.3 & 5.3 \\
10261$-$5055 & 3000 & $-$0.5 & 0.2 & 1.0$\times$10$^{-9}$ & 4 & $\leq-$1.0 & $\leq$ 0.0 & 5.8 & 5.8 \\
11081$-$4203 & 3000 & $-$0.5 & 1.6 & 5.0$\times$10$^{-8}$ &8{\textsuperscript{$\dagger$}}& 1.3 & 1.1 & 5.8 & 5.8 \\
%11525$-$5027 & 3300 & $-$0.5 & ... & ... & ... & 0.9 & ... & ... & ...       \\
%12377$-$6102{\textsuperscript{*}} & ... & $-$0.5 & ... & ... & 20 & ... & ... & ... & ... \\
14266$-$4211 & 2900 & $-$0.5 & 0.2 & 5.0$\times$10$^{-8}$ & 9 & $\leq$ 0.0 & 0.2 & $\leq$ 5.3 & $\leq$ 5.3   \\
14337$-$6215{\textsuperscript{*}} & 3300 & $-$0.5 & 0.2 & 5.0$\times$10$^{-8}$ & 20 & 2.4$^{1}$ & 2.4$^{1}$ & ... & ... \\
15193$+$3132 & 2800 & $-$0.5 & 1.6 & 1.0$\times$10$^{-9}$ & 3 & $\leq$ 0.0 & $\leq$ 0.0 & $\leq$ 5.8 & $\leq$ 5.8   \\
15211$-$4254 & 3300{\textsuperscript{**}} & $-$0.5 & 1.6 & 1.0$\times$10$^{-9}$ & 11 & 2.3 & 2.3 & 5.8 & 5.8\\
15255$+$1944 & 2900 & $-$0.5 & 0.2 & 5.0$\times$10$^{-7}$ & 7 & 1.0 & 0.9 & 6.0 & 6.0   \\
15576$-$1212 & 3000 & $-$0.5 & 0.2 & 1.0$\times$10$^{-8}$ & 10 & 1.1 & 1.2 & 5.7 & 5.7     \\
16030$-$5156{\textsuperscript{*}} & 3000 & $-$0.5 & 0.2 & 1.0$\times$10$^{-8}$ &10{\textsuperscript{$\dagger$}}& 1.5 & 1.5 & ... & ... \\
16037$+$4218 & 2900 & $-$0.5 & 1.2 & 1.0$\times$10$^{-8}$ & 4 & $\leq-$1.0 & $\leq$ 0.0 & 6.3 & 6.3 \\
16260$+$3454 & 3300{\textsuperscript{**}} & $-$0.5 & 0.2 & 1.0$\times$10$^{-9}$ & 12 & 2.7 & 2.6 & 5.4 & 5.4\\
17034$-$1024 & 3300{\textsuperscript{**}} & $-$0.5 &0.8&1.0$\times$10$^{-8}$&8{\textsuperscript{$\dagger$}}&$\leq$ 0.0&$\leq$ 0.0& 6.2 & 6.2 \\
%18025$-$2113{\textsuperscript{1}} & ... & $-$0.5 & ... & ... & 11 & ... & ... & ... & ... \\
%18057$-$2616{\textsuperscript{2}} & 3000 & $-$0.5 & ... & ... & 19 & ... & ... & ... & ...    \\
18413$+$1354 & 3300{\textsuperscript{**}}& $-$0.5 & 1.2 & 1.0$\times$10$^{-8}$ & 15 & 1.8 & 1.7&$\leq$ 5.3& $\leq$ 5.3 \\
18429$-$1721 & 3000 & $-$0.5 & 0.2 & 1.0$\times$10$^{-8}$ & 7 &  1.2 & 1.1 & 5.8 & 5.8       \\
%19059$-$2219{\textsuperscript{**}} & 3000 & $-$0.5 & ... & ... & 13 & ... & ... & ... & ...     \\
19129$+$2803 & 3300{\textsuperscript{**}} & $-$0.5 & 0.2 & 1.0$\times$10$^{-8}$ &11$^{\dagger}$& 3.1$^{1}$ & 3.1$^{1}$ & 5.5 & 5.5 \\
19361$-$1658 & 3000 & $-$0.5 & 0.2 & 1.0$\times$10$^{-9}$ & 8 & 1.9 & 2.0 & 5.8 & 5.8    \\
%19426$+$4342{\textsuperscript{**}} & 3000 & $-$0.5 & ... & ... & 9 & ... & ... & ... & ...     \\
20052$+$0554{\textsuperscript{*}} & 3300 & $-$0.5 & 0.2 & 1.0$\times$10$^{-7}$ & 16 & 2.6 & 2.3 & ... & ...\\
%20077$-$0625{\textsuperscript{**}} & 3000 & $-$0.5 & ... & ... & 12 & ... & ... & ... & ...   \\
20343$-$3020 & 3000 & $-$0.5 & 1.2 & 1.0$\times$10$^{-9}$ & 8 & $\leq-$1.0 & $\leq$ 0.0 & 6.0 & 6.0       \\
RU Cyg & 3000 & $-$0.5 & 1.6 & 1.0$\times$10$^{-7}$ & 12$^{\dagger}$ & 2.0 & 1.7 & 6.2 & 6.2  \\
SV Cas & 3000 & $-$0.5 & 1.6 & 1.0$\times$10$^{-7}$ & 12$^{\dagger}$ & 3.5 & 3.3 & 6.3 & 6.3        \\
R Cen$^{2}$ & 3000 & $-$0.5 & 1.6 & 1.0$\times$10$^{-7}$ & 5$^{\dagger}$ & 4.3 & 4.0 & ... & ... \\
\hline
\end{tabular}
\end{center}
\textsuperscript{$\dagger$}\footnotesize{These stars only display a single peak in the 1612 MHz OH maser, so we adopt an average velocity in the abundance analysis exploring OH expansion velocities of other stars with similar variability periods.} \\
\textsuperscript{*}\footnotesize{The S/N at Ca I 6463 \AA~ is very low to derive any Ca abundance estimate.}\\ 
\textsuperscript{**}\footnotesize{The best fitting T$_{eff}$} in the Ca I 6463 \AA~ spectral region is cooler (3000 K) than the one around the Li I 6708 \AA~ line.\\
%\textsuperscript{**}\footnotesize{Optical counterpart too red.} \\
%\textsuperscript{1}\footnotesize{Possible double lined spectroscopic binary.} \\
%\textsuperscript{2}\footnotesize{The Li I line has a P cygni-type profile.} \\
\textsuperscript{1}\footnotesize{The line is resolved in two components (circumstellar and photospheric) and the abundance obtained corresponds to the photospheric one.} \\
\textsuperscript{2}\footnotesize{R Cen was not observed in the Ca I line region.} \\
%\textsuperscript{5}\footnotesize{The best fitting T$_{eff}$} in the Ca I 6462.6 \AA~ spectral region is cooler (3000 K) than the one around the Li I 6708 \AA~ line.\\
%\tablefoot{We have checked the uncertainties in the Li and Ca derived abundances of the sample stars. We have made small changes in the model atmosphere parameters (${\Delta}T_{eff}$ = $\pm$100 K, ${\Delta}$log $g$ = $\pm$0.5, ${\Delta}Z$ = $\pm$0.2, ${\Delta}$t = $\pm$0.5 km s$^{-1}$, ${\Delta}{\beta}$ = $\pm$0.2, ${\Delta}log$($\dot{M}/M_{\odot}$yr$^{-1}$) = $\pm$0.5, ${\Delta}v_{exp}$(OH) = $\pm$5 km s$^{-1}$, $\Delta$FWHM = $\pm$50 m$\AA$), which result in error bars of $\pm$0.3 and $\pm$0.2 dex in the hydrostatic and pseudo-dynamic Li abundances respectively.In the Ca case, the formal uncertainties due to changes in the model atmosphere parameters are $\pm$0.5 dex both in the hydrostatic and pseudo-dynamic abundances.}
\end{table*}

The atmospheric and wind parameters as well as the Li and Ca abundances (or 
upper limits) from the best fits to the observed spectra are shown in Table 
\ref{table_abundances}. The new Li abundances determined from the extended 
models are very similar to those obtained with the hydrostatic models (see 
Table \ref{table_abundances}). A maximum difference, between the hydrostatic 
and dynamical abundances ($\Delta$(log$\varepsilon$(Li))$_{_{static-dynamic}}$), 
of $+$0.3 dex is found for IRAS 20052$+$0554 and R Cen, while an average 
difference of $+$0.18 dex is found in our entire AGB sample. This indicates that 
the Li content in these stars is not strongly affected by the presence of a 
circumstellar envelope. It is to be noted here that the latter number does not 
consider the four stars (IRAS 05098$-$6422, IRAS 10261$-$5055, IRAS 16037$+$4218 
and IRAS 20343$-$3020) for which we get more conservative pseudo-dynamic Li upper 
limits ($\leq$0.0 dex) than from the hydrostatic models ($\leq$-1.0 dex). In  
addition, the Ca abundances from the hydrostatic and pseudo-dynamical models 
are practically identical and the presence of a circumstellar envelope does not 
affect at all the derived Ca abundances. 

We have estimated the uncertainties in the derived Li and Ca abundances for the sample stars. For this, we have made small changes in the atmosphere parameters ${\Delta}T_{eff}$ = $\pm$100 K, 
${\Delta}$log $g$ = $\pm$0.5, ${\Delta}Z$ = $\pm$0.2, ${\Delta}$t = $\pm$0.5 km s$^{-1}$ and ${\Delta}FWHM$ = $\pm$50 m$\AA$ for the hydrostatic models, and also in the wind
parameters ${\Delta}{\beta}$ = $\pm$0.2, ${\Delta}log$($\dot{M}/M_{\odot}$yr$^{-1}$) = 
$\pm$0.5 and ${\Delta}v_{exp}$(OH) = $\pm$5 km s$^{-1}$ for the pseudo-dynamic models. These small changes result in Li formal errors of $\pm$0.3 and $\pm$0.2 dex for the hydrostatic and 
pseudo-dynamic abundances, respectively, while the estimated Ca formal uncertainties are $\pm$0.5 dex for both the hydrostatic and pseudo-dynamic abundances.

%
%______________________________________________________________

\section{Non-LTE effects on the Li I and Ca I lines}

Due to the fact that the classical hydrostatic and our pseudo-dynamical synthetic spectra are constructed by considering a local thermodynamic equilibrium (LTE) treatment in the 
(extended) stellar atmosphere, we have explored the possible non-LTE (NLTE) effects on the 6708 \AA~Li I and 6463 \AA~Ca I resonance lines in order to clarify the sign and magnitude of the corrections to be applied to the hydrostatic and pseudo-dynamic Li and Ca abundances.

The NLTE radiative transfer calculations for the 6708 \AA~Li I line
were performed using the {\tt MULTI} code \citep{carlsson86, 
carlsson92} with the same Li atom model as in \cite{osorio11}. This Li atom model includes quantum mechanical calculations of electron 
and hydrogen collisional excitation as well as charge exchange with 
hydrogen \citep[see more details in][]{osorio11}. The calculations were 
performed in the same grid of MARCS models used in \cite{osorio16} 
and the models are hydrostatic. For this study we focused on 
atmospheric models with log $g$ $=-0.5$, [Fe/H] = 0.0 and five 
different effective temperatures, T$_{eff}$ = {2500, 2600, 2700, 
3300 and 3400} K. For the three coolest models, the NLTE abundance 
corrections $\Delta$(log$\varepsilon$(Li))$_{_{NLTE-LTE}}$ are $\gtrapprox$ $+$0.2 dex, reaching $+$0.3 dex at log$\varepsilon$(Li)=0.0. The warmer model atmospheres display smaller NLTE abundance corrections of $\Delta$(log$\varepsilon$(Li))$_{_{NLTE-LTE}}$ $\sim$ 0.0 dex at 
log$\varepsilon$(Li) $\sim$ $+$2.0 and $\Delta$(log$\varepsilon$(Li))$_{_{NLTE-LTE}}$ $\sim$ $+$0.1 dex around log$\varepsilon$(Li) $\geq$ 3.0 and log$\varepsilon$(Li) $\leq$ 0.5. Our updated NLTE Li calculations thus confirm the Li-rich character of our massive O-rich AGB sample stars and that the adoption of LTE in Li-rich AGB stars is likely to result in an underestimation of the Li abundances \citep[e.g.][]{kiselman95, abia99}.

For the NLTE calculations of the 6463 \AA~Ca I line, the {\tt MULTI} code was used with the same Ca atom model as in \cite{osorio18}. This Ca atom model includes also updated data for electron and hydrogen collisional excitation and charge exchange with hydrogen. The calculations were performed in 
two atmospheric hydrostatic models with T$_{eff}$/log$g$/[Fe/H] = 
3000/$-$0.5/0.0 and 3300/$-$0.5/0.0 for which we found positive NLTE abundance corrections of $\sim$ $+$0.06 and $+$0.02 dex, respectively. In short, our NLTE Ca calculations show that the use of LTE in massive O-rich AGB stars would translate into a slight underestimation of the real Ca abundances (see Subsection 6.2).

%
%_____________________________________________________________________

\section{Discussion}

\subsection{Lithium}
Our Li results in massive Galactic AGB stars, including a circumstellar component 
in the analysis, do not reflect a dramatic change in the derived abundances, 
contrary to our previous findings on the Rb abundances in these stars 
\citep{zamora14,perez-mesa17}. The Rb abundances obtained with pseudo-dynamical 
models are much lower (sometimes even by 1-2 dex) than the hydrostatic ones, being 
strongly affected by the presence of a circumstellar envelope. We have made 
several tests by changing the wind parameters (mass loss rate $\dot{M}$, 
parameter $\beta$ and the terminal velocity $v_{exp}$(OH)) in the models; e.g. 
not fixing $\dot{M}$ and  $\beta$ or by assuming the wind model parameters from the best 
fits of the Rb I 7800 $\AA$ line \citep{perez-mesa17} but the Li abundances
remain very similar (within 0.1 dex). As we have mentioned before, for consistency, we have fixed $\dot{M}$ and $\beta$ to those values obtained by \cite{perez-mesa17} from the Rb spectral fits (when available) because the synthetic spectra around 7800 $\AA$ are more sensitive to 
variations in the model wind parameters than the Li I 6708\AA~spectral region. Our finding of the circumstellar effects being not so important for the Li I 6708\AA~line could be somehow surprising because the atomic parameters
(e.g. excitation potential) of the Li I 6708 $\AA$ and Rb I 7800 $\AA$
resonance lines are quite similar. The lower Li abundance as compared with Rb, 
and therefore lower Li I column-density in the circumstellar envelope, however,
likely explains the small abundance differences obtained between the pseudo-dynamical and
hydrostatic models. In addition, other factors such as the molecular blends in each wavelength range and
the line depth formation could influence the
different sensitivity to the circumstellar effects between Rb and Li. In particular, the line depth formation is extremely important because the velocity field could change the $\tau$-scale of the lines \citep[see][]{nowotny10}.

We have compared our new Li abundances with up-to-date solar metallicity 
massive AGB nucleosynthesis models with very different prescriptions for 
mass loss and convection: (i) ATON models \citep{ventura18} with the 
\cite{bloecker95} recipe for mass loss and the full spectrum of turbulence 
convective mixing \citep[FST; e.g.][]{mazzitelli99}; (ii) Monash models 
\citep{karakaslugaro16} with the \cite{vassiliadis93} mass-loss prescription 
and the mixing length theory of convection \cite[MLT;][]{bohm58}; (iii) 
FRUITY\footnote{FUll-Network Repository of Updated Isotopic Tables and Yields: 
http://fruity.oa-abruzzo.inaf.it/.} models \citep{cristallo15} with a
pulsationally-driven mass-loss rate \citep[see][]{straniero06} and the 
MLT of convection but under the formulae from \cite{straniero06}; and 
(iv) NUGrid/MESA models \citep{ritter18} with the mass-loss formula 
from \cite{bloecker95} and assuming convective boundary mixing 
\cite[CBM; e.g.][]{ritter18}. 

%\textbf{Figure \ref{Li_Ca_M} (\textit{left panel}) 
%shows the comparation between the massive AGB nucelosynthesis models and
%the obtained Li abundances in our sample.}

The evolution of Li in massive HBB-AGB stars is strongly affected by several 
stellar parameters such as progenitor mass, metallicity, mass loss and
convection  model \citep[see e.g.][]{mazzitelli99, vanraai12}. During the AGB
phase, the mass  loss and the treatment of the convection are the most important
factors in the  determination of the duration of HBB and the variation of the
surface chemistry  during the AGB phase. For example, i) the massive AGB
nucleosynthesis ATON models  show strong Li abundance oscillations (by orders of
magnitude) on timescales as  short as $\sim$10$^{4}$ years \citep[see
e.g.][]{mazzitelli99} and there may be negligible Li in the envelope for a
significant (at least 20\%) period of time; and ii) the Li minima and the
duration of the Li-rich phase in the Monash models are less deep and
longer, respectively, than in the ATON models \citep[see][for more
details]{garcia-hernandez13}. This complex theoretical evolution of the Li
abundance implies that the Li abundances distribution derived from the
spectroscopic observations (e.g. the exact progenitor mass and
evolutionary status are not known) can only be analyzed in a statistical way
\citep[][]{garcia-hernandez07}.

Regarding the peak surface Li abundances during the AGB, the ATON models predict that it goes from log$\varepsilon$(Li) = 3.8 dex for M = 3.5 M$_{\odot}$ to 4.3 dex for M = 6.0 and 7.5 M$_{\odot}$, while in the Monash models it changes from log$\varepsilon$(Li) = 3.8 dex for M = 4.25 M$_{\odot}$ to log$\varepsilon$(Li) = 4.4 dex for M = 8 M$_{\odot}$. In the NuGrid/MESA models, Li production at Z = 0.02 is only predicted for M = 6 and 7 M$_{\odot}$ with a peak surface Li abundance of 2.9 and 3.7 dex, respectively. However, the FRUITY models do not predict production of Li at all, which is at odds with the Li overabundances observed in massive AGB stars in the Galaxy \citep[e.g.][]{garcia-hernandez07,garcia-hernandez13}, the Magellanic Clouds \citep[e.g.][]{plez93, smith95, 
garcia-hernandez09} and the Li-detected O-rich AGB star in the dwarf galaxy IC 1613 \citep[e.g.][]{menzies15}.

The pseudo-dynamic Li abundances obtained from our spectra are between $\sim$0.0
and 4.0 dex; with eight non Li-rich (log$\varepsilon$(Li) $<$ 0.5 dex), twenty
Li-rich (0.5 $\leq$ log$\varepsilon$(Li) $\leq$ 3.2 dex) and two super Li-rich
(log$\varepsilon$(Li) $>$ 3.2 dex) stars. Their great similarity with the
hydrostatic Li abundances (and the relatively small positive NLTE corrections;
Subsect. 5) means that the conclusions reached by
\cite{garcia-hernandez07,garcia-hernandez13} are unchanged and will not be
repeated here. In short, the Li-rich and super Li-rich character of the massive
AGB stars in our sample confirm that they experience strong HBB
\citep{garcia-hernandez07,garcia-hernandez13}. This is in good agreement with
the predictions from AGB nucleosynthesis models like the ATON, Monash and
NuGrid/MESA, but in strong contrast with the FRUITY AGB models, which do not
predict strong HBB and Li production in solar metallicity massive AGB stars.

%\begin{figure*}
%\centering
%\includegraphics[width=9.1cm,height=6.7cm,angle=0]{Li_M_stars.png}
%\includegraphics[width=9.1cm,height=6.7cm,angle=0]{Ca_M_stars.png}
%\caption{\textbf{Predictions of stellar mass vs. log$\varepsilon$(Li)
%(\textit{left panel}) and log$\varepsilon$(Ca) (\textit{right panel})
%from the ATON, Monash, NUGrid/MESA and FRUITY nucleosynthesis models.
%The shaded regions mark the range of the log$\varepsilon$(Li) and
%log$\varepsilon$(Ca) obtained in our sample with pseudo-dynamical
%models.}}
%\label{Li_Ca_M}
%\end{figure*}

%
%______________________________________________________________

\subsection{Calcium}

This is the first work in which the Ca abundances have been obtained for  a
sample of massive AGB stars. Figure \ref{comparisons_Ca} shows that the 6463
$\AA$ Ca I line is not sensitive to changes in the stellar (T$_{eff}$) and wind
($\dot{M}$, $\beta$ and $v_{exp}$(OH)) parameters. Thus, we adopted the wind
parameters from the Rb fits (when possible) or the Li fits, which are more
sensitive to variations of the wind parameters. The hydrostatic and
pseudo-dynamic abundances of Ca obtained from our spectra are identical; so the
Ca abundances are not affected at all by the presence  of a circumstellar
envelope. The Ca abundances obtained are in the range log  $\varepsilon(Ca)$ =
5.3 - 6.3 dex. The theoretical AGB nucleosynthesis models predict an important
production of some Ca isotopes like the radioactive $^{41}$Ca but no significant change in
the total Ca abundance (see Section 1). Note that here we consider log
$\varepsilon(Ca)$ = 6.31  dex as the solar abundance in the photosphere
\citep{grevesse07}. While ATON models \citep{ventura18} do not include Ca,
the Monash models predict Ca abundances in the range log $\varepsilon(Ca)$ =
6.31 - 6.35 dex for solar metallicity; i.e. a 12\% increase at most
relatively to the initial value used in these models of 6.29 dex. In the same
way, the FRUITY models predict solar Ca abundances  for Z = 0.014. In the
NuGrid/MESA models, Ca at Z = 0.01 is predicted to vary from  log
$\varepsilon(Ca)$ = 6.11 to 6.18 dex in the range of M = 3 -7 solar masses, 
while for Z = 0.02 the Ca abundances are between 6.44 and 6.48 dex. 

%\textbf{In
%Figure \ref{Li_Ca_M} (\textit{right panel}) are shown the theoretical
%predictions of Ca from the massive AGB nucleosynthesis models.}

\begin{figure}
\centering
\includegraphics[width=7.0cm,height=9.0cm,angle=90]{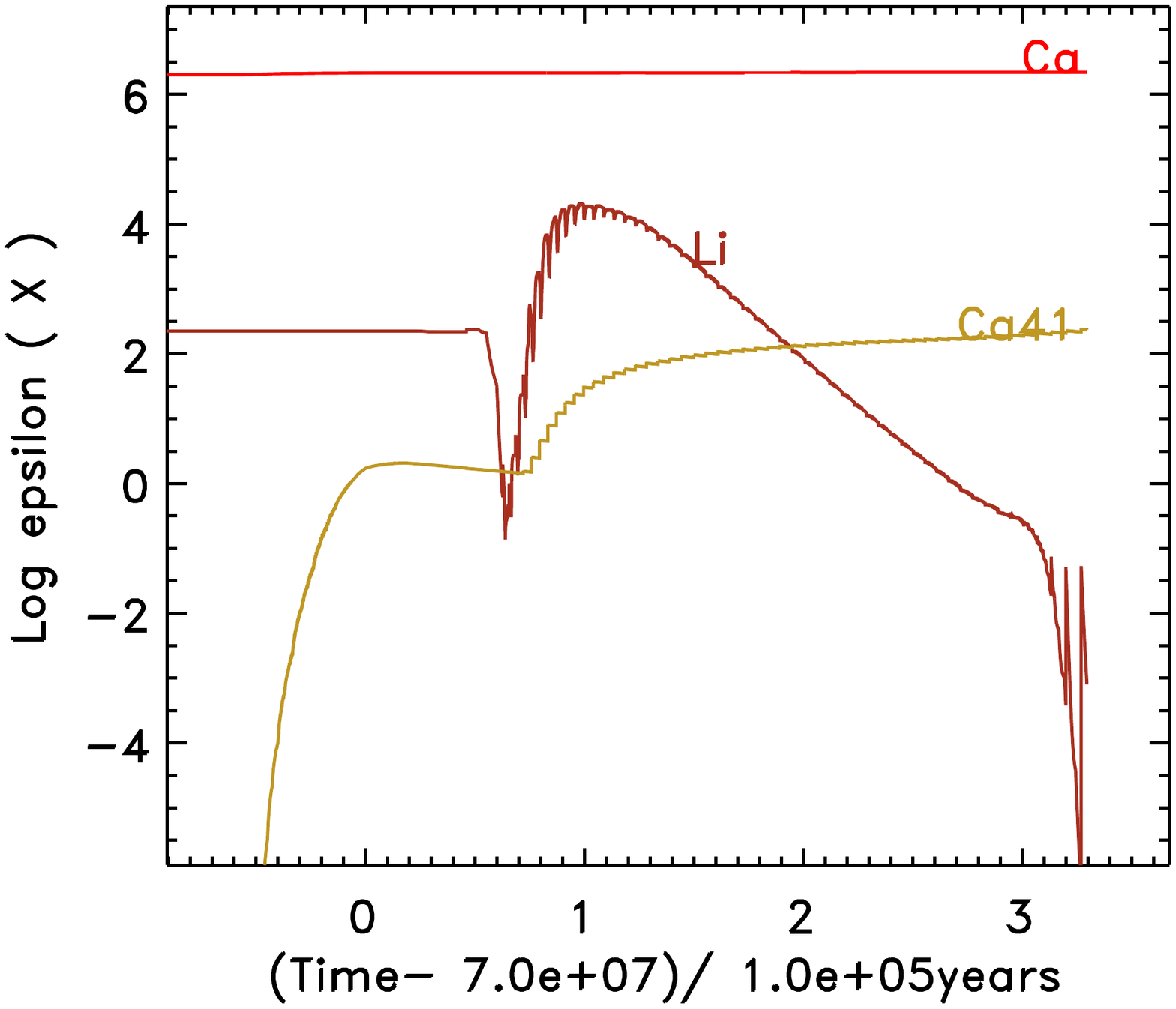}
\caption{Evolution of Li, Ca and $^{41}$Ca versus time for the
6 M$_{\odot}$ model of solar metallicity from \cite{karakaslugaro16}.}
\label{m6z014}
\end{figure}

In Figure \ref{m6z014} we show the evolution of Li, Ca and the radioactive
isotope $^{41}$Ca as a function of time for the 6 M$_{\odot}$ model of solar
metallicity from \cite{karakaslugaro16}. This figure shows that the 
$^{41}$Ca abundance increases as a consequence of nucleosynthesis and
mixing during the TP-AGB phase although the first increase
of the $^{41}$Ca occurred during the second dredge-up, after core helium
burning. We can also see that the total elemental Ca abundance is, however,
unchanged. The 6 M$_{\odot}$ model has TDU and HBB as described in 
\cite{karakaslugaro16}, remains oxygen rich as a consequence of HBB, and as shown by
Figure \ref{m6z014} the model becomes Li-rich, where the $\log
\varepsilon$(Li) exceeds 3 for $\sim$ 80,000 years.

We find that most (20) sample stars display Ca abundance values $\sim$ 0.5 - 0.6
dex lower than the adopted Ca solar abundance of log $\varepsilon(Ca)$ = 6.31
dex. In spite of the fact that we cannot completely discard that some of our
sample stars could be indeed slightly metal poor\footnote{It is to be noted,
however, that most sample stars are expected to be of solar metallicity
\citep[see a detailed discussion on this in][]{garcia-hernandez07}.}, their Ca
abundances can be considered as nearly solar when taking into account our
estimated Ca abundance errors ($\sim$ 0.5 dex) and the possible NLTE effects
(see Subsect. 5). This is consistent with the predictions from the available
\textit{s}-process nucleosynthesis models for solar metallicity massive AGB
stars, as mentioned above. 

However, a minority (5) of the sample stars seem to show a significant Ca 
depletion ($-0.8$ to $-1.0$ dex). We explored if the derived Ca  abundances are
correlated with the Li content, the wind parameters (mass  loss $\dot{M}$, beta
parameter $\beta$ and terminal velocity $v_{exp}$(OH)) or other observational
information such as variability periods, near-IR colors and IR excess but our
search proved to be negative with the possible exception of the near-IR colors
and the IR excess (see below). We have identified three possibilities (i.e.
missed opacities in  the stellar atmosphere models, Ca depletion into dust and
line weakening  phenomena) in order to understand their apparent (and
unexpected) Ca  depletion and that are enumerated below:

\begin{figure*}
\centering
\includegraphics[width=9.1cm,height=6.7cm,angle=0]{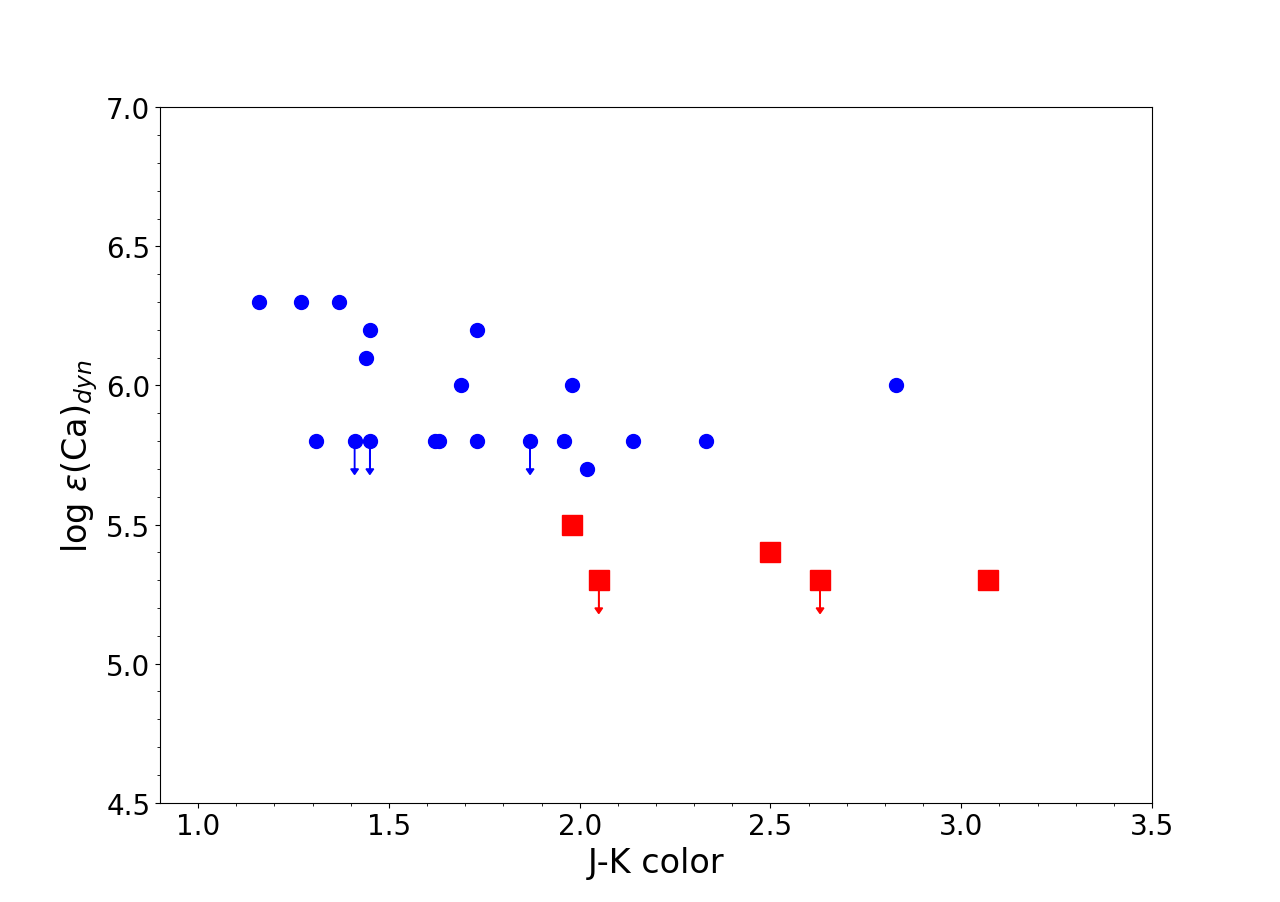}
\includegraphics[width=9.1cm,height=6.7cm,angle=0]{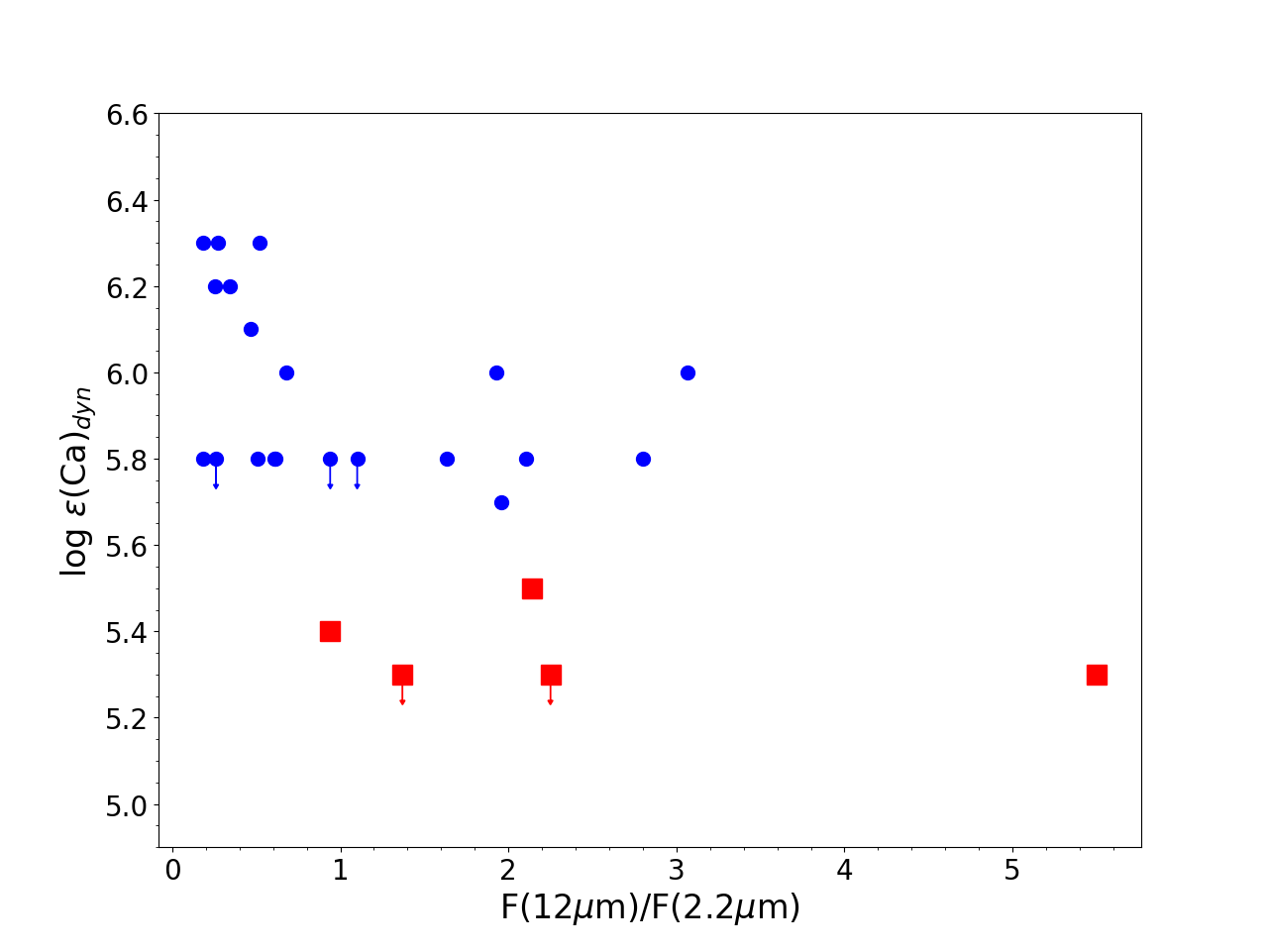}
\caption{The Ca abundances derived with pseudo-dynamical models against  the J-K
colors (left panel) and infrared excesses  R = F(12$\mu$m)/F(2.2$\mu$m) (right
panel). The 5 stars with larger Ca depletion are shown by red squares and upper
limits are marked with vertical arrows.}
\label{Ca_vs_color}
\end{figure*} 

i) The low Ca abundances in our sample stars could be due to missed opacities in
the stellar atmosphere models. Although the Ca spectral region (mainly dominated
by the TiO molecule) is generally well modelled by us,
\citet{garcia-hernandez07,garcia-hernandez09} have reported the presence of
strong and yet unidentified molecular bands in several spectral regions in the
optical spectra of massive Galactic and extragalactic O-rich AGB stars, which
suggest the presence of other opacity contributors not yet considered in the
model atmospheres and in the construction of synthetic optical spectra for O-rich
AGB stars.

ii) Although the condensation of inorganic dust grains in the winds of evolved
stars is still poorly understood, the observed Ca underabundances may be also
due to the fact that Ca in our sample of stars could be depleted into dust
\citep[see e.g.][for a review]{lodders99}. Figure \ref{Ca_vs_color}
plots the Ca pseudo-dynamical abundaces against the 2MASS J-K colors and the
infrared excesses R = F(12$\mu$m)/F(2.2$\mu$m)\footnote{The infrared excess R
probes the presence of circumstellar material emitting at 12 $\mu$m with respect
to the stellar continuum at 2.2 $\mu$m \citep[see e.g.][]{jorissen93}.} (with
fluxes at 2.2 and 12 $\mu$m from 2MASS and IRAS, respectively) in our sample
stars. Curiously, the 5 stars with a significant Ca depletion are the redder
ones (they display, on average, higher J-K colors) and most of them display a
significant infrared excess, suggesting that they could be among the more
evolved and/or dusty stars in our sample. However, the number of stars is still
low and Galactic massive AGB stars are known to display a large photometric
variability in the near-IR and mid-IR ranges \citep[see
e.g.][]{garcia-hernandez07b}. \cite{dellagli14} studied the alumina dust
(the amorphous state of Al$_{2}$O$_{3}$) production in O-rich circumstellar shells, which is  expected
to be fairly abundant in the winds of the more massive and  O-rich AGB stars. By
coupling AGB stellar nucleosynthesis and dust formation, the predicted production
of alumina dust implies an important decrease (see below) in the abundance of
gaseous Al in the AGB wind. The high fraction of gaseous Al condensed in
Al$_{2}$O$_{3}$ (especially in their more massive AGB models) implies that the
gaseous Al is expected to be underabundant in the more massive HBB-AGB stars;
something that is in good agreement with the only estimate of the Al content in
massive HBB-AGB stars to date\footnote{\cite{mcsaveney07} measured $\log
\varepsilon$(Al) = 5.5 dex in the truly massive HBB-AGB star HV 2576 in the LMC 
(Z = -0.3 dex), while the Al solar abundance is 6.4 dex \citep{grevesse07}. The
amount of gaseous Al depleted into dust is 0.6 dex, in good agreement with the
\cite{dellagli14} AGB models that include dust formation.}; i.e. the Al content
measured in a confirmed massive HBB-AGB in the Large Magellanic Cloud
\citep[see][for more details]{dellagli14}. Thus, similarly to the Al case, the
gaseous Ca in massive Galactic O-rich AGB stars could be depleted into dust. For
example, \cite{tielens90} proposed a dust condensation sequence in O-rich
circumstellar regions, in which the formation of calcium-rich silicates such as
augite (Ca$_{2}$Al$_{2}$SiO$_{7}$), diopside (CaMgSi$_{2}$O$_{6}$) and anothite
(CaAl$_{2}$Si$_{2}$O$_{6}$) would be expected 
\citep[see also][and references therein]{lodders99}. However, \cite{speck00} suggested
that the dust evolutionary path is different for the AGB and red supergiant (RSG)
stars; although both condensation sequences eventually would lead to similar dust
types. Basically, the main difference between the RSG and AGB dust condensation
sequences is that the RSG stars experience an evolutionary phase in which
aluminium- and calcium-rich silicates condensation takes place, while this
evolutionary phase is apparently not seen in the AGB stars. \cite{speck00}
classified the spectra of a sample of AGB and RSG stars into several groups
according to the observed appearance of the amorphous silicates infrared (IR)
features around 10 $\mu$m. They found that the RSG IR spectra are better
reproduced when calcium-rich silicates are considered, while the AGB stars are
well reproduced with amorphous silicates only. Unfortunately, we have only three
stars (RU Cyg, IRAS 07304$-$2032 and IRAS 15193$+$3139) in common with
\cite{speck00} and they are not among the most Ca-poor stars in our sample.
\cite{speck00} classified their spectra as \textit{silicate A} (RU Cyg) and 
\textit{silicate B} (IRAS 07304$-$2032 and IRAS 15193$+$3139) AGB types; all of
them with no clear signs for calcium-rich silicates. Additional N-band IR 
spectroscopic observations of confirmed Galactic AGB O-rich stars (especially for
those stars with significant Ca depletion) would be desirable in order to clarify
if their 10 $\mu$m amorphous silicates dust features could be better fitted by
the inclusion of Ca-rich silicates. Ca-rich stardust grains from AGB stars have
been recovered from meteorites, also belonging to the  Group II population that
probably originated from HBB-AGB stars \citep{lugaro17}. Both hibonite grains
\citep[e.g.][]{nittler08}, as well as Ca-rich silicates  \citep[e.g.][]{nguyen04,
vollmer09} have been reported.

iii) Finally, line weakening phenomena could be another possibility to explain
the lack of Ca in these sample stars. \cite{humphreys74} studied some
high-luminosity M-type supergiants that show veiling of the absorption metallic
lines; the veiling effect was found to be most pronounced in the near-IR than in
the blue spectral regions. \cite{humphreys74} proposed that the peculiar energy
distributions of these stars and the veiling of the absorption lines may be
explained by a combination of free-bound emission ($\lambda<$ 1.6 $\mu$m) and
free-free emission ($\lambda>$ 1.6 $\mu$m) from electron-neutral H interactions
arising in the extended atmosphere around the star plus the surrounding
circumstellar shell of dust grains. Such line weakening phenomenon, as observed
in these peculiar M-supergiants, could be also present in similar M-type
long-period variables (more than 260 days) dusty stars such as our sample stars.
We have looked for additional absorption metallic lines in our Li and Ca spectral
regions (e.g. the 6469 $\AA$ Fe I and 6484 $\AA$ Ni I lines) in order to check if
line weakening phenomena are affecting other metallic lines. Unfortunately, our
optical spectra are severely dominated by the TiO molecule and no metallic lines
are detected. We thus cannot confirm or discard if the lack of Ca in our minority
stars is because of possible line weakening phenomena affecting the optical
spectra of massive Galactic O-rich AGB stars.

%
%______________________________________________________________

\section{Conclusions}

We have reported new hydrostatic and pseudo-dynamical abundances of Li and Ca
from the 6708 $\AA$ Li I and 6573 $\AA$ Ca I lines, respectively, in a complete
sample of massive Galactic O-rich AGB stars by using a modified version of the
spectral synthesis code \textit{Turbospectrum}, which considers the presence of
a circumstellar envelope with a radial wind. 

The new Li abundances from pseudo-dynamical models are very similar to those
obtained  from the hydrostatic models (the average difference is 0.18 dex), while
they are identical for Ca. This indicates that the determination of the Li and Ca
abundances in massive  O-rich AGB stars is not strongly affected by the presence
of a circumstellar envelope.  Indeed, we found that the the Li I and Ca I line
profiles are not very sensitive to  variations of the wind ($\dot{M}$, $\beta$
and $v_{exp}$(OH)) parameters.

The new pseudo-dynamic abundances of Li (30 stars) confirm the Li-rich (and super
Li-rich in some stars) character of our sample stars and the strong activation of
the HBB process in massive Galactic AGB stars. This is in good agreement with the
theoretical predictions from the most recent AGB nucleosynthesis models such as
ATON, Monash and NuGrid/MESA, but at odds with the FRUITY database, which
predicts no Li production by HBB in massive AGB stars at solar metallicity.

For the first time we have obtained Ca abundances in a sample of massive
Galactic  AGB stars. Most of them (20) display nearly solar Ca abundances; within
the estimated errors and/or considering possible NLTE effects. Their abundances
are thus consistent with the predictions from the \textit{s}-process
nucleosynthesis models for massive AGB stars at solar metallicity. For example,
such models predict some production of the radioactive $^{41}$Ca isotope but no
change in the total Ca abundance. A minority of stars (5) show a significant Ca
depletion (by $\sim$ -0.8 $-$ -1.0 dex). Possible explanations to explain their
apparent and unexpected Ca depletion could be missed opacities in the stellar 
atmosphere models and/or Ca depletion into dust as well as line weakening
phenomena.

%
%____________________________________________________________________________

\begin{acknowledgements}
The authors thank Flavia Dell'Agli for providing the Li peak abundances
from the ATON models and Umberto Battino and Ashley Tattersall for providing 
the Li and Ca abundances from the NuGrid/MESA models. This work is based on 
observations at the 4.2 m William Herschel Telescope operated on the island 
of La Palma by the Isaac Newton Group in the Spanish Observatorio del Roque 
de Los Muchachos of the Instituto de Astrof\'isica de Canarias. Also based 
on observations with the ESO 3.6 m telescope at La Silla Observatory (Chile). 
V.P.M. acknowledges the financial support from the Spanish Ministry of Economy 
and Competitiveness (MINECO) under the 2011 Severo Ochoa Program MINECO 
SEV$-$2011$-$0187. V.P.M., O.Z., D.A.G.H., T.M. and A.M. acknowledge support 
provided by MINECO under grant AYA$-$2017$-$88254-P. M.L. is a Momentum 
(“Lend\"ulet-2014” Programme) project leader of the Hungarian Academy of 
Sciences. This paper made use of the IAC Supercomputing facility HTCondor 
(http://research.cs.wisc.edu/htcondor/), partly financed by the Ministry of 
Economy and Competitiveness with FEDER funds, code IACA13$-$3E$-$2493.
\end{acknowledgements}

\begin{appendix}
\section{Complete sample}\label{Append_sample}
\begin{figure*}
   \centering
   \includegraphics[width=9.1cm,height=6.5cm,angle=0]{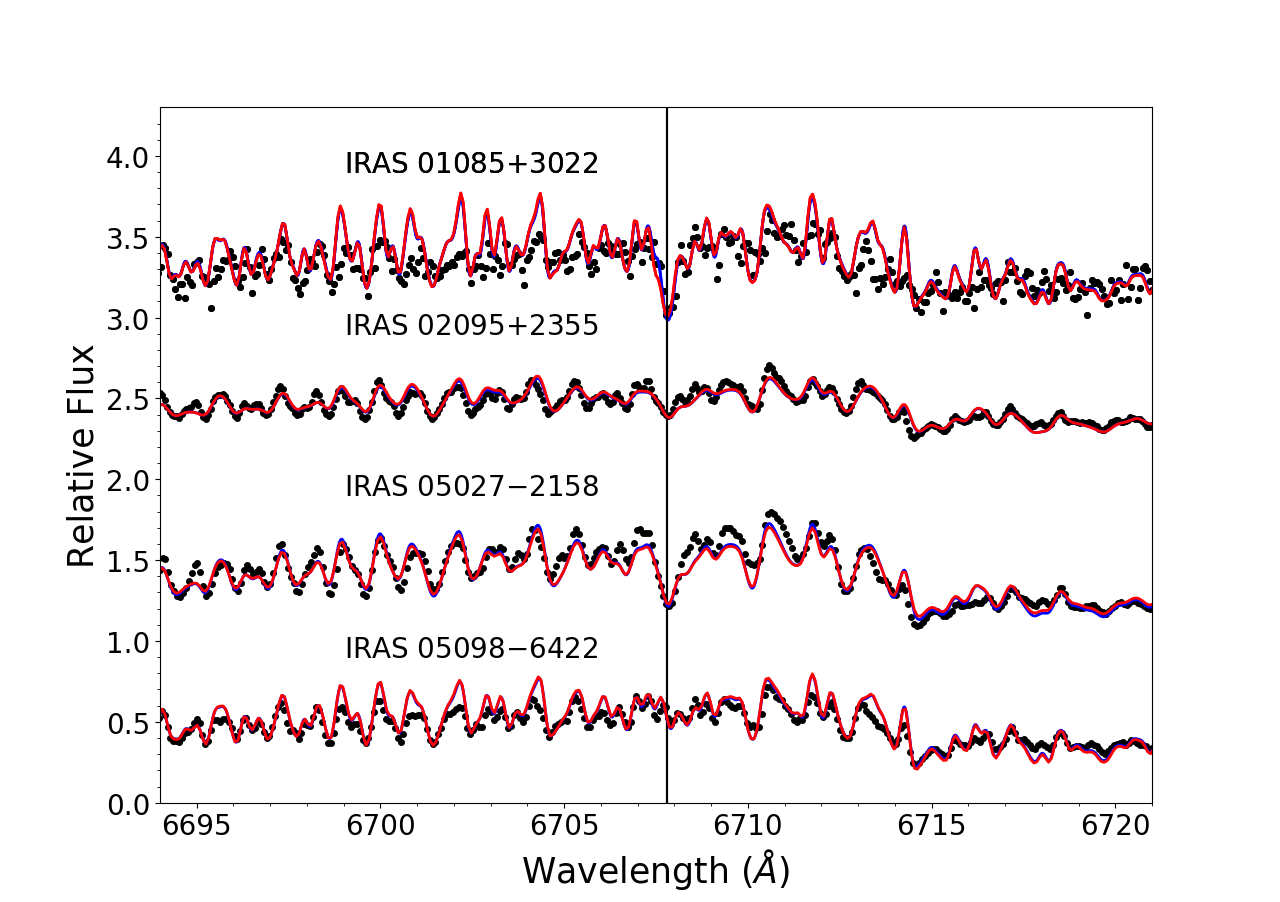}
   \includegraphics[width=9.1cm,height=6.5cm,angle=0]{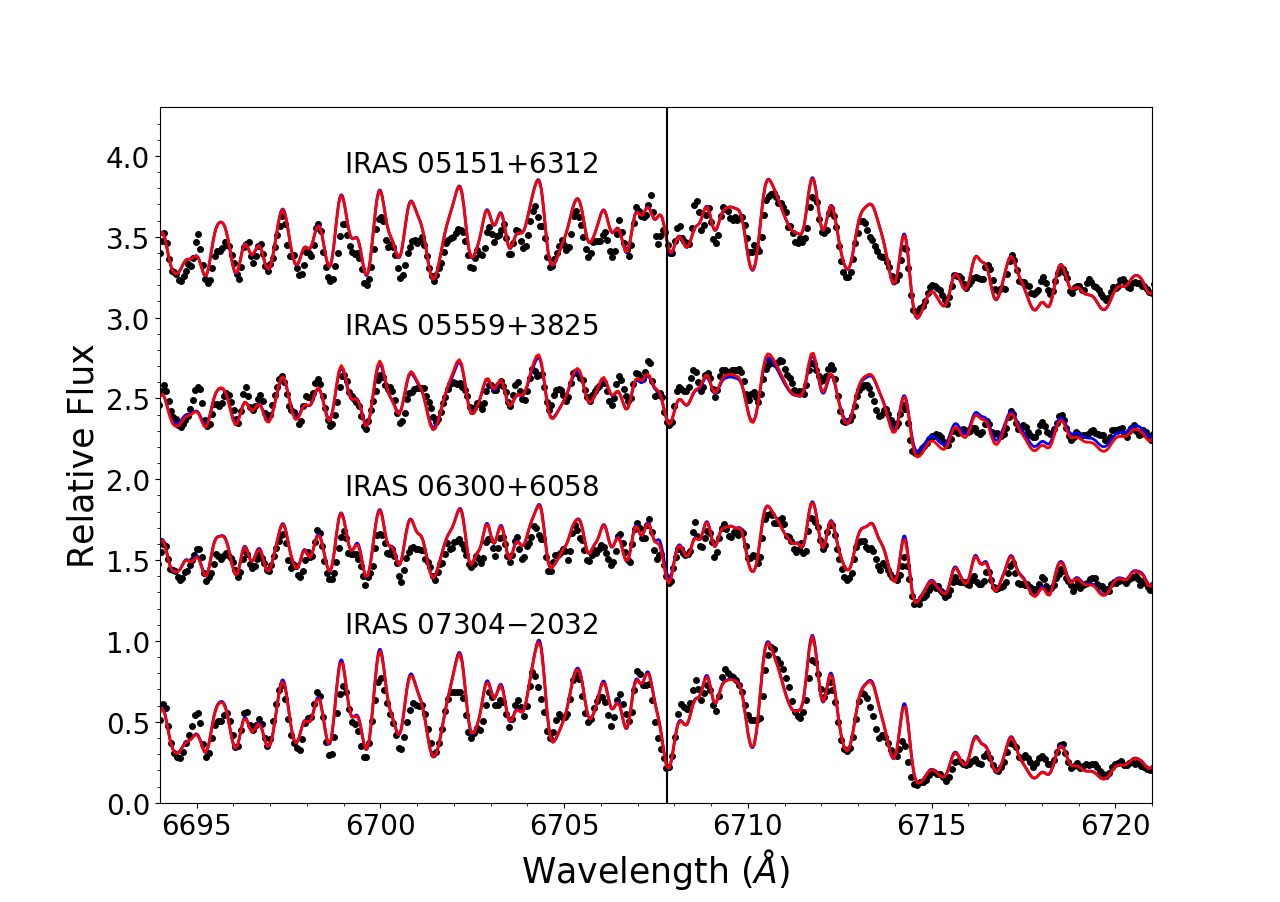}
   \includegraphics[width=9.1cm,height=6.5cm,angle=0]{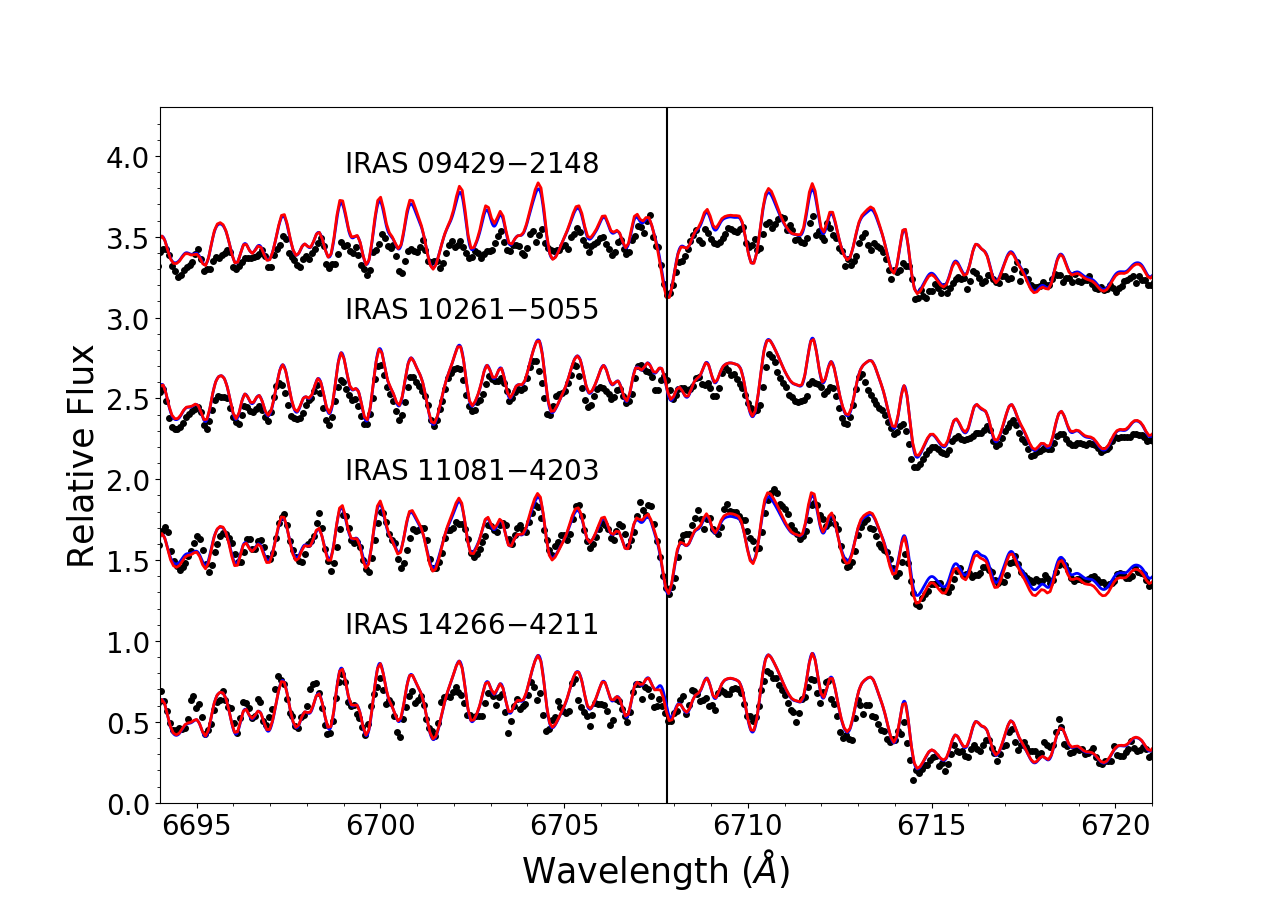}
   \includegraphics[width=9.1cm,height=6.5cm,angle=0]{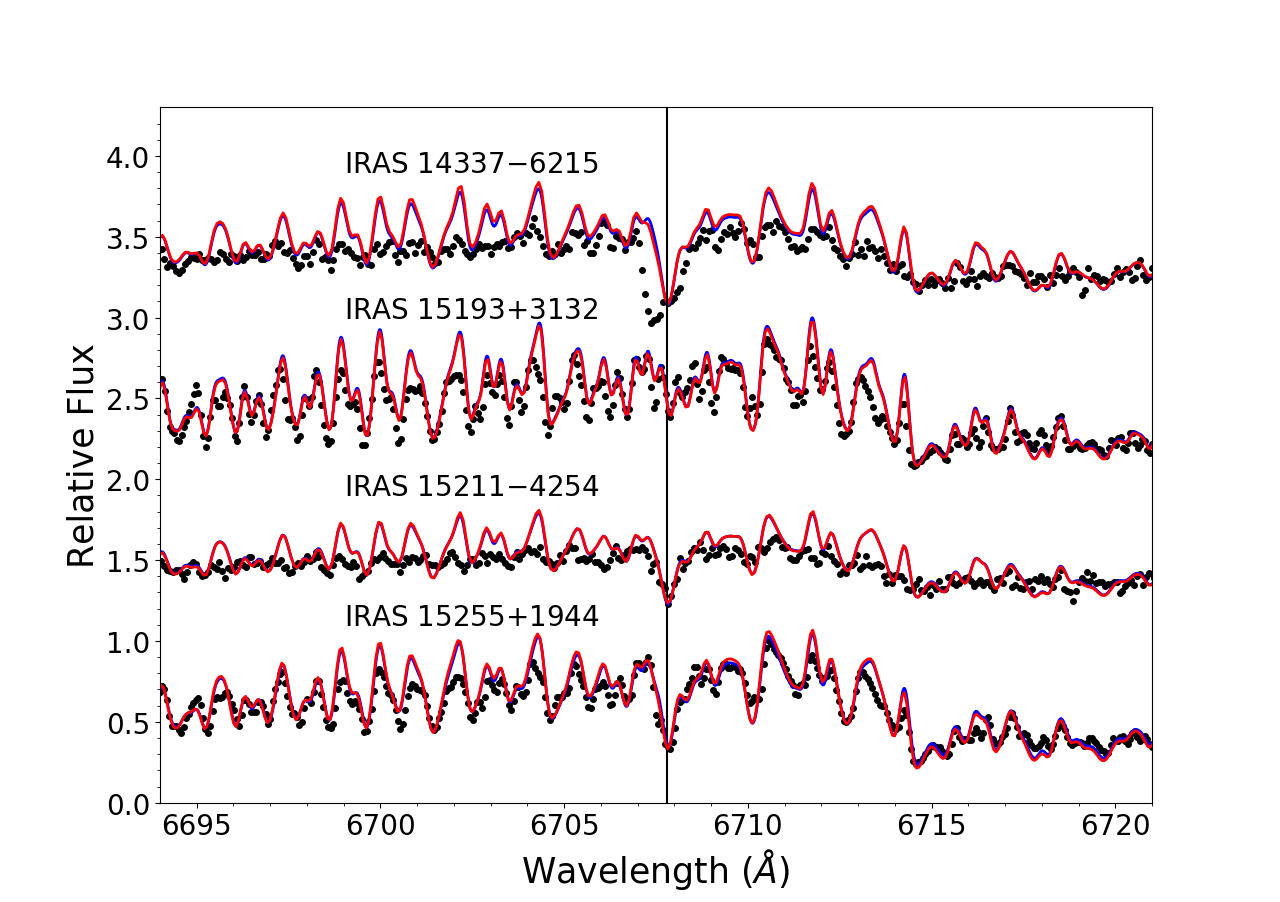}
   \includegraphics[width=9.1cm,height=6.5cm,angle=0]{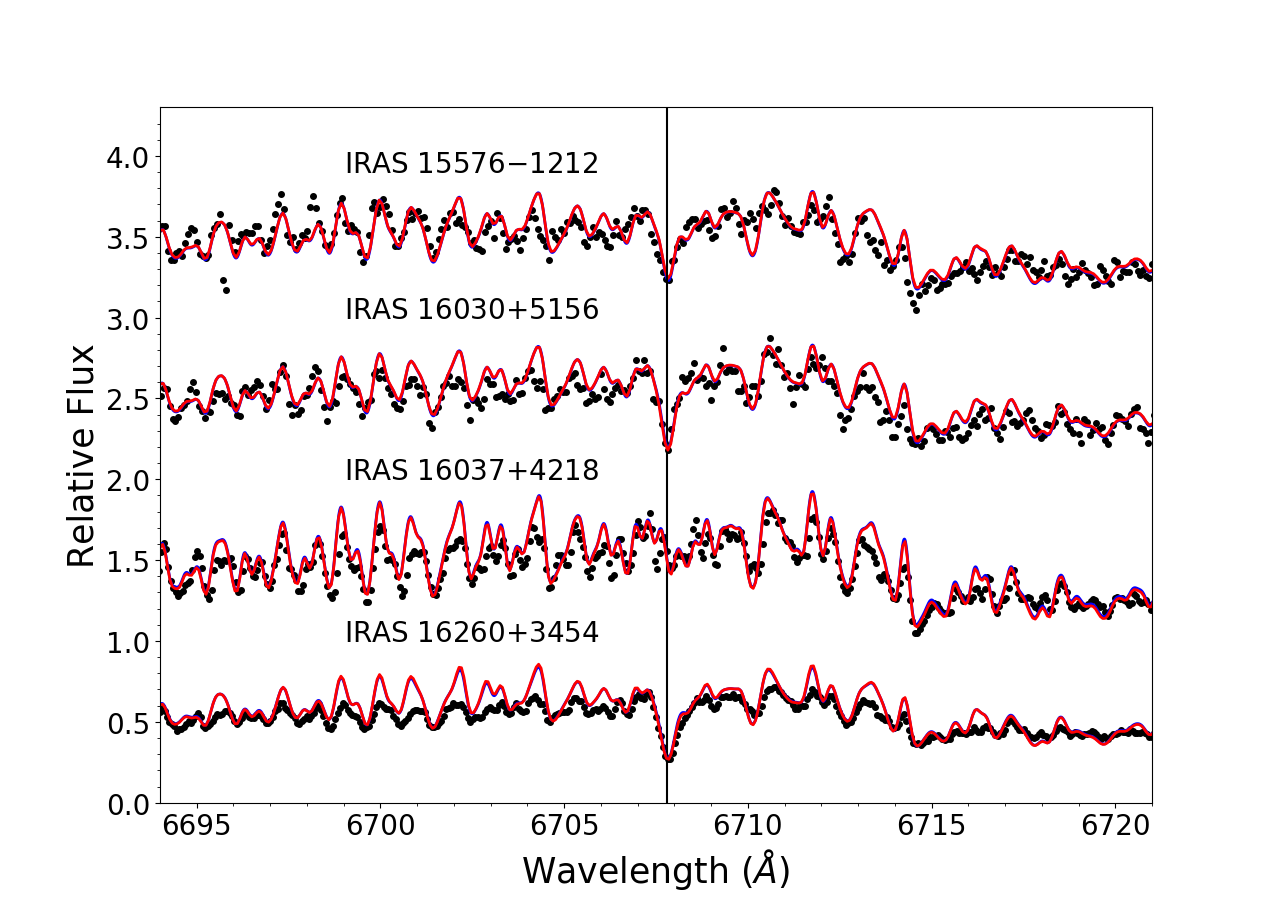}
   \includegraphics[width=9.1cm,height=6.5cm,angle=0]{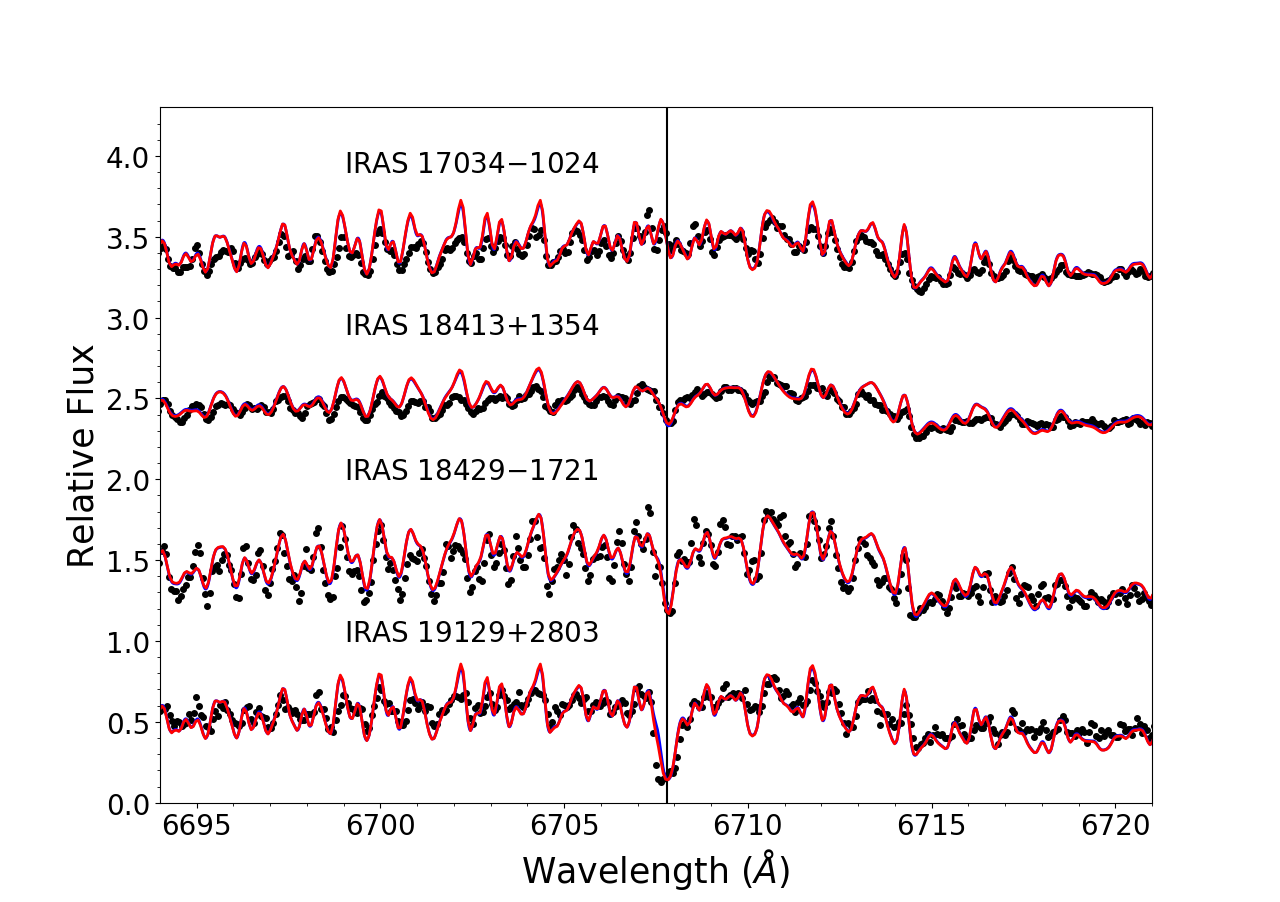}
   \includegraphics[width=9.1cm,height=6.5cm,angle=0]{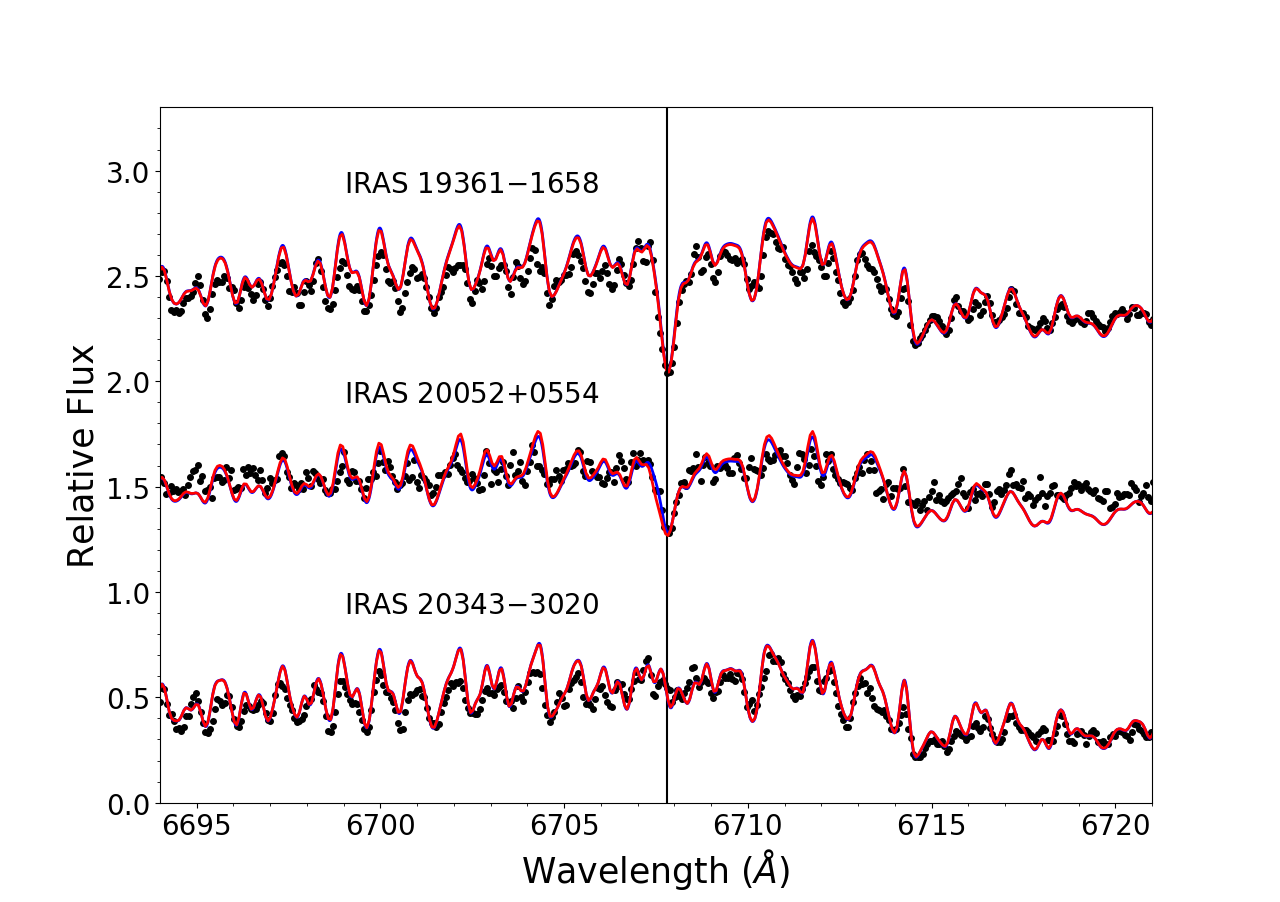}
   \includegraphics[width=9.1cm,height=6.5cm,angle=0]{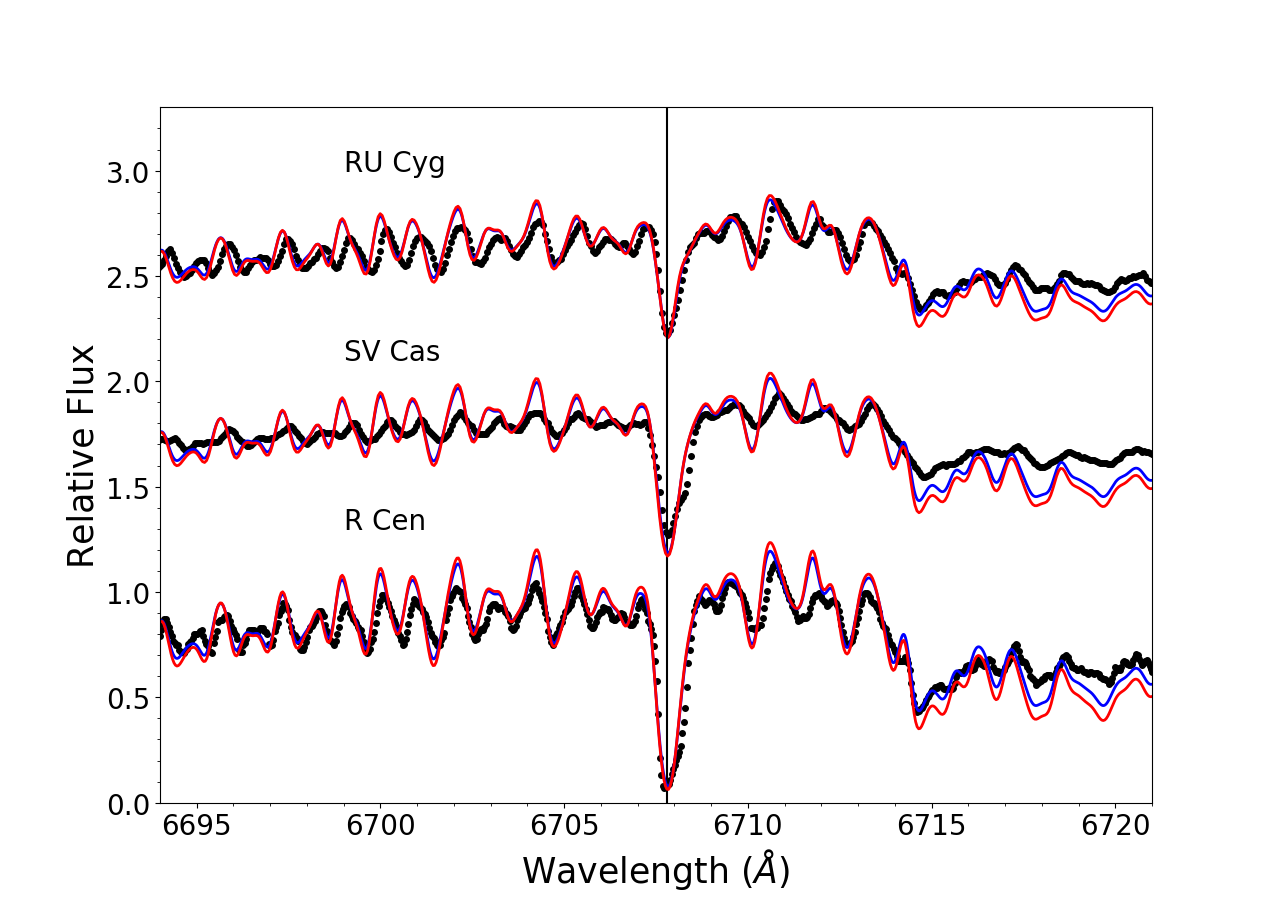}
   \caption{Observed spectra (black dots), best hydrostatic (blue lines) and pseudo-dynamical (red lines) fits of our sample of AGB stars in the regions of 6708 $\AA$ Li I line. The parameters of the best fit model atmospheres are indicated in Table \ref{table_abundances}. The plots are displayed in increasing R.A. order.}
   \label{Li_sample}
\end{figure*}

\begin{figure*}
   \centering
   \includegraphics[width=9.1cm,height=6.5cm,angle=0]{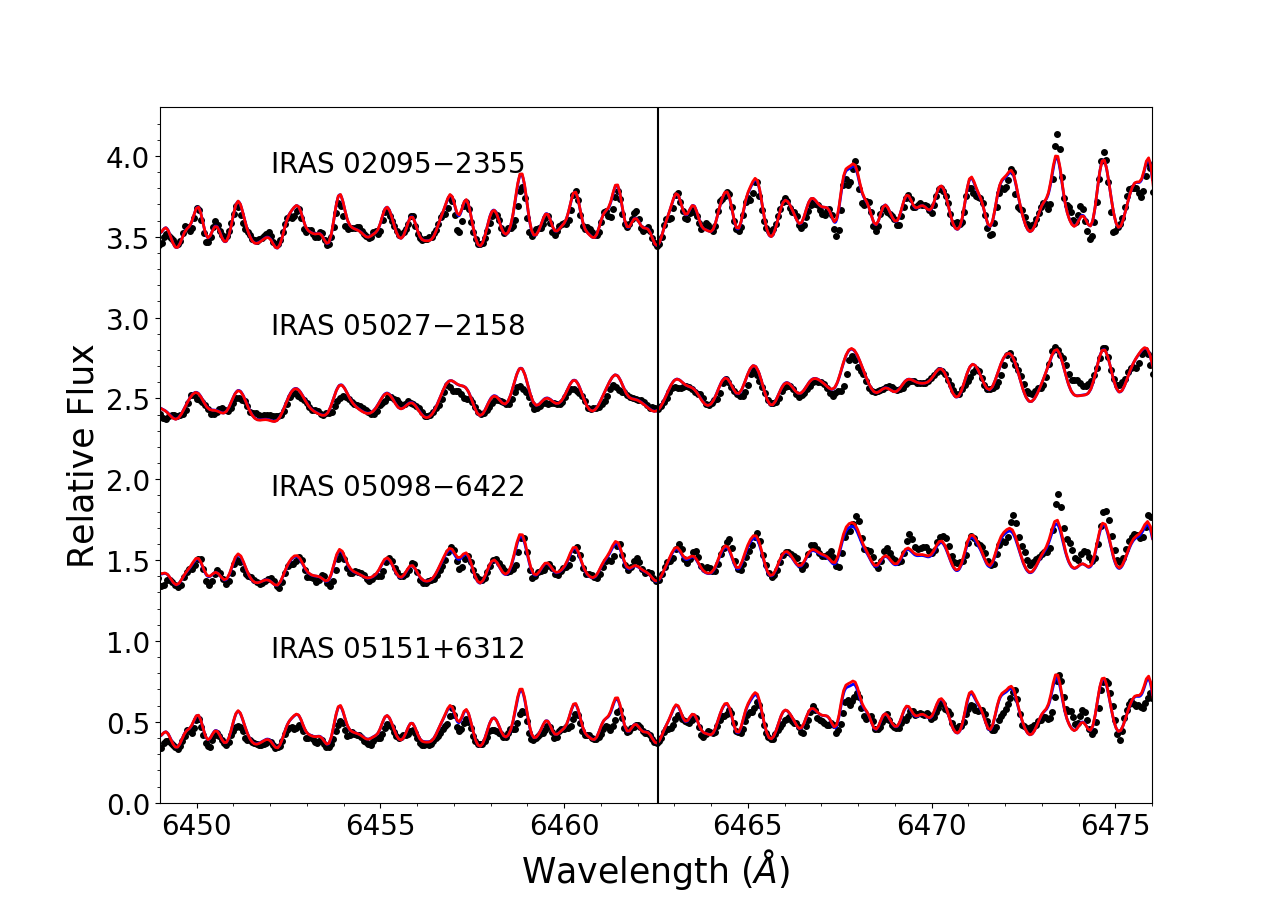}
   \includegraphics[width=9.1cm,height=6.5cm,angle=0]{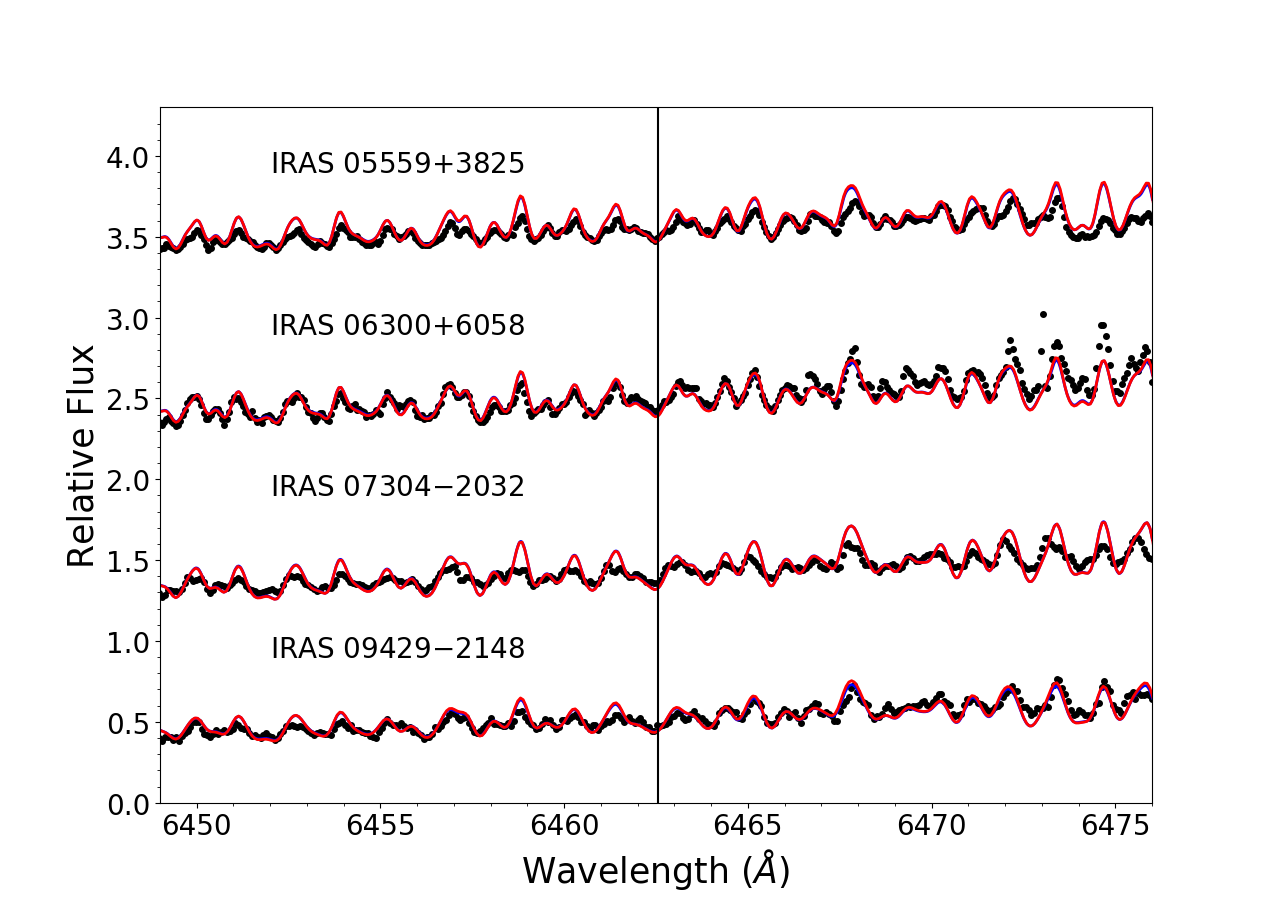}
   \includegraphics[width=9.1cm,height=6.5cm,angle=0]{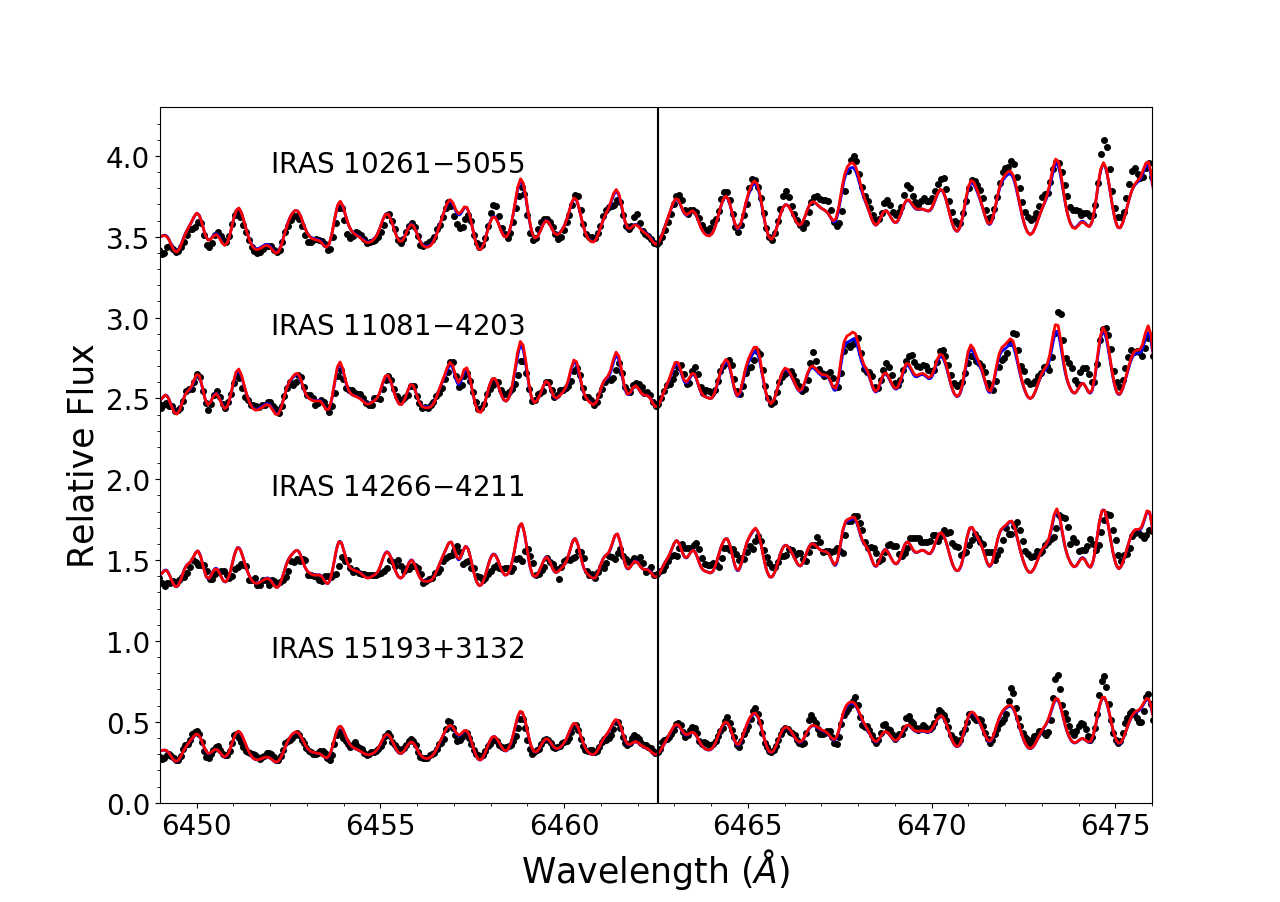}
   \includegraphics[width=9.1cm,height=6.5cm,angle=0]{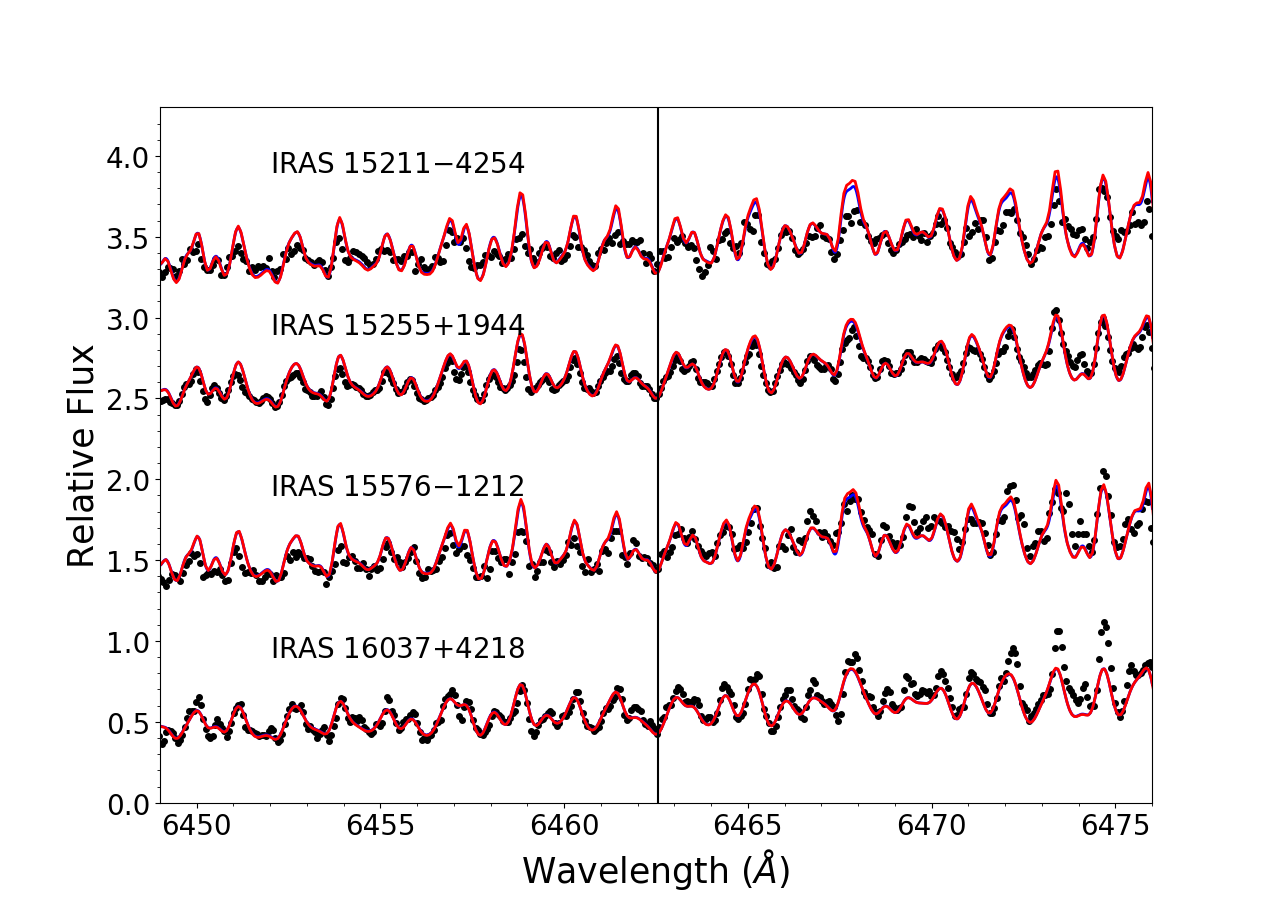}
   \includegraphics[width=9.1cm,height=6.5cm,angle=0]{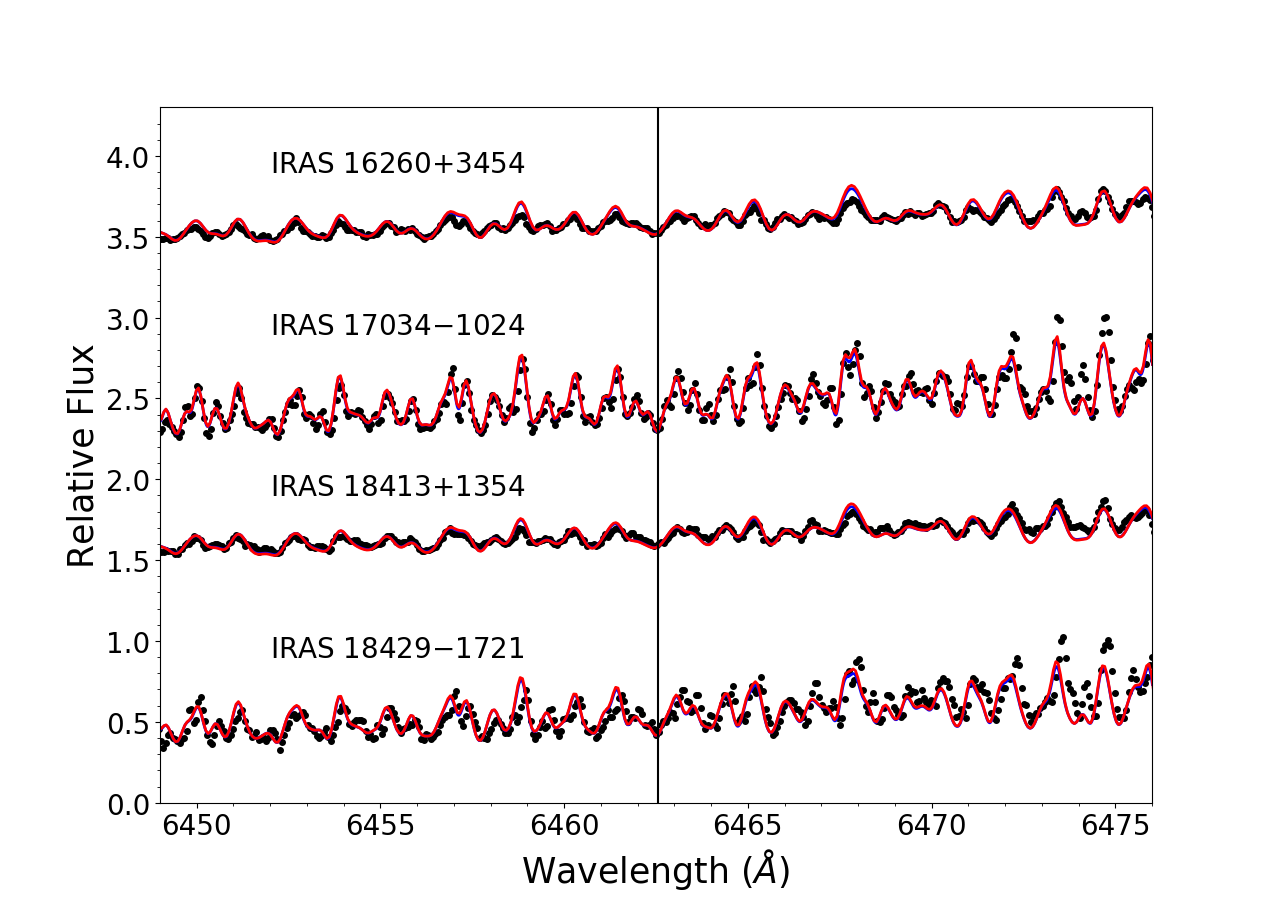}
   \includegraphics[width=9.1cm,height=6.5cm,angle=0]{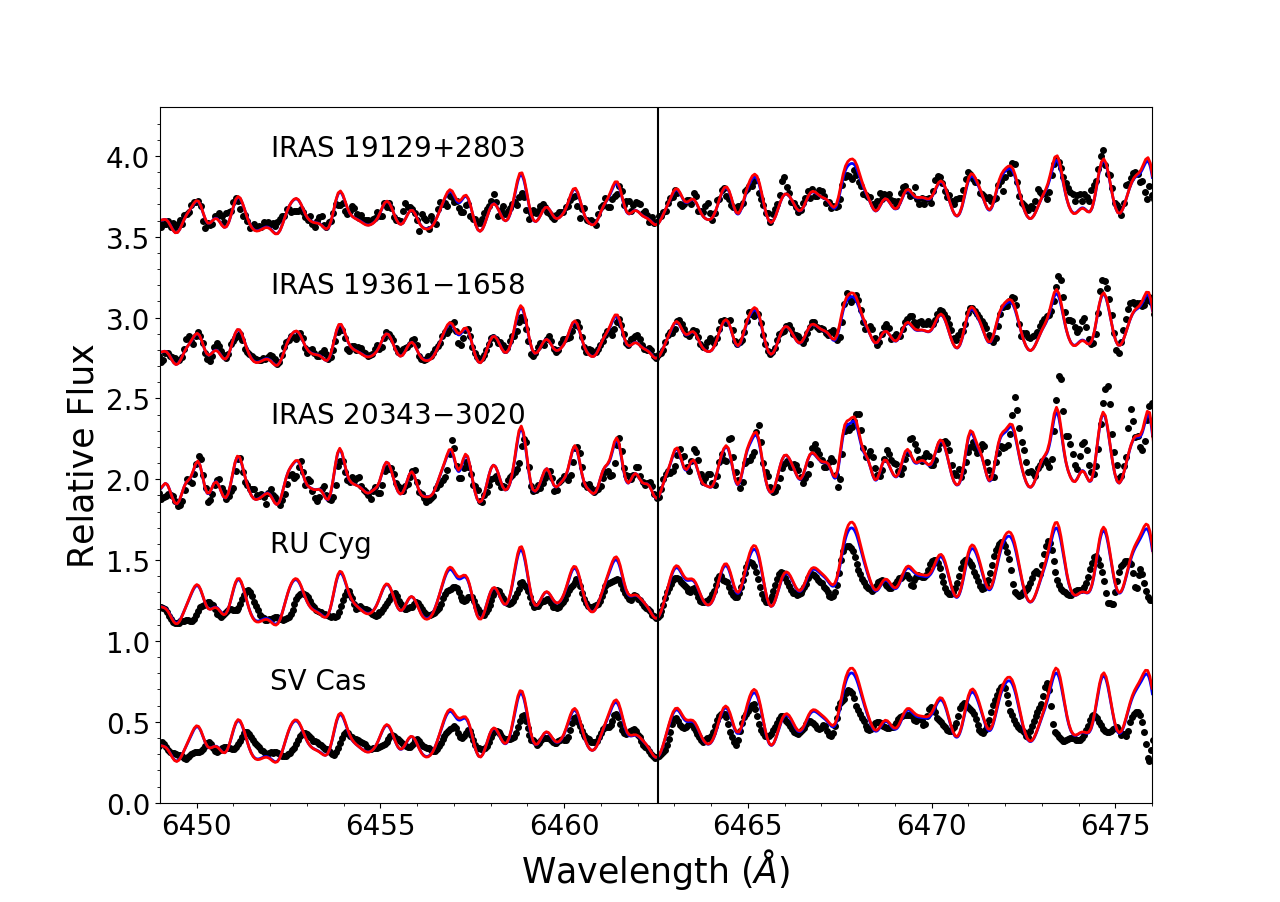}
   \caption{Observed spectra (black dots), best hydrostatic (blue lines) and pseudo-dynamical (red lines) fits of our sample of AGB stars in the regions of 6463 $\AA$ Ca I line. The parameters of the best fit model atmospheres are indicated in Table \ref{table_abundances}. The plots are displayed in increasing R.A. order.}
   \label{Ca_sample}
\end{figure*}

The best fits of the 6708 $\AA$ Li I and 6463 $\AA$ Ca I spectral regions of our
sample of massive AGB stars are displayed in Figure \ref{Li_sample} and
\ref{Ca_sample} respectevely. The pseudo-dynamical models are similar to 
the hydrostatic ones, and  reproduce properly the Li and Ca regions, which
means that the Li and Ca line profiles are not  strongly affected by the 
presence of a circumstellar envelope and a radial wind. 

\end{appendix}

% WARNING
%-------------------------------------------------------------------
% Please note that we have included the references to the file aa.dem in
% order to compile it, but we ask you to:
%
% - use BibTeX with the regular commands:
%   \bibliographystyle{aa} % style aa.bst
%   \bibliography{Yourfile} % your references Yourfile.bib
%
% - join the .bib files when you upload your source files
%-------------------------------------------------------------------

\end{document}